\documentclass[%
 reprint,
 showpacs,
 showkeys,
 preprintnumbers,
 amsmath,amssymb,
 aps,
 pra,
  longbibliography,
 ]{revtex4-1}

\usepackage[breaklinks=true,colorlinks=true,anchorcolor=blue,citecolor=blue,filecolor=blue,menucolor=blue,pagecolor=blue,urlcolor=blue,linkcolor=blue]{hyperref}
\usepackage{graphicx}
\usepackage{xcolor}
\usepackage[left]{eurosym}
\usepackage{url}

\begin{document}

\title{An Apology for Money}

\author{Karl Svozil}
\affiliation{Institute of Theoretical Physics, Vienna
    University of Technology, Wiedner Hauptstra\ss e 8-10/136, A-1040
    Vienna, Austria}
\email{svozil@tuwien.ac.at} \homepage[]{http://tph.tuwien.ac.at/~svozil}

\date{\today}

\begin{abstract}
This review is about the convenience, the benefits, as well as the destructive capacities of money. It deals with various aspects of money creation, with its value, and its appropriation. All sorts of money tend to get corrupted by eventually creating too much of them. In the long run, this renders money worthless and deprives people holding it. This misuse of money creation is inevitable and should come as no surprise. Abusive money creation comes in various forms. In the present fiat money system ``suspended in free thought'' and sustained merely by our belief in and our conditioning to it, money is conveniently created out of ``thin air''  by excessive government spending and speculative credit creation. Alas, any too tight money supply could ruin an economy by inviting all sorts of unfriendly takeovers, including wars or competition. Therefore the ambivalence of money as benefactor and destroyer should be accepted as destiny.
\end{abstract}

\maketitle

\if01

\NeedsTeXFormat{LaTeX2e}[1996/06/01]

\documentclass[final]{blurb}

\usepackage{rotating}
\usepackage{floatpag}
\rotfloatpagestyle{empty}
\usepackage[left]{eurosym}
\usepackage{lettrine}
\RequirePackage[scaled=.90]{helvet}
\RequirePackage{courier}
\usepackage[T1]{fontenc}
 \usepackage[urw-garamond]{mathdesign}

\renewcommand{\baselinestretch}{1.15}

\usepackage[breaklinks=true,colorlinks=true,anchorcolor=blue,citecolor=blue,filecolor=blue,menucolor=blue,pagecolor=blue,urlcolor=blue,linkcolor=blue]{hyperref}
\usepackage{doi}
\usepackage{natbib,url}
\bibliographystyle{newapa-doi-url}


 \usepackage{makeidx}
 \makeindex




  \hyphenation{line-break line-breaks docu-ment triangle cambridge amsthdoc
    cambridgemods baseline-skip author authors cambridgestyle en-vir-on-ment polar}

  \setcounter{tocdepth}{2}




\begin{document}
\citeindextrue
\sloppy

  \title[A Very Brief Guide to the Ambivalence of Money\\
           As a Benefactor and Destroyer]
    {An Apology for Money}

\author{Karl Rudolf Svozil
\vskip 10 true cm
Edition Funzel}

  \frontmatter
  \maketitle
  \tableofcontents

  \mainmatter







\fi

\subsection{Preface}

\begin{quote}
No one is more a slave than the man who thinks himself free while he is not.
({\em Johann Wolfgang von Goethe}, in {\it Elective Affinities}.)
\end{quote}

\if01

{{A}}{fter} ridding themselves from a nuclear power plant just after its completion,
some concerned Austrians used to joke,
``where does all the electricity come from?
Well, of course it comes from the socket!''
So, by analogy, one could ask,
``Where does all our money come from?''
and rightly respond,
``from the teller machine,''
or more generally,
``from the bank!''
Whereas  in the electrical case the answer appears polemic and foolish,
in the money case it is exactly correct: money is ``produced''
by the bank; ``out of thin air,'' or, stated differently, ``out of nothing.''

Why, then, do we have so little money if it can be created so easily?
Because the essence of money is its scarcity,
or at least the impression of scarcity.\index{scarcity}
For if money was abundant and readily available everywhere, or if anybody could have access to it effortlessly ``on demand,''
it would get inflated and be worth nothing.
This, in a sentence, refutes monetary systems without interest or privilege.

So money appears to be both producible ``out of thin air,''
and at the same time valuable.
How to achieve that apparent contradiction is straightforward:
first,
some ``authorities'' have to be entrusted with the creation of money;
and second,
some provisions for limits to its appropriation should render, at least the impression of, scarcity.

How did I start to think about money?
\fi

{{A}}{ couple} of years ago, as we were having lunch
at the cafeteria,
a colleague from the physics department  passed me a seemingly simple question, ``how exactly is money created?''
After some {\it ad hoc}
attempts of ``explaining'' this to him and to myself, I ended up staring at my half-eaten salad in bewilderment.

I almost gasped. I could not tell. I was at a miserable loss when it came to money creation.  I felt humiliated and perplexed.

Now
I am convinced that I can answer that question.
Yet I had to absorb some uncomforting facts.

Indeed, at some point I felt like the main character  in Carpenter's 1988 movie {\em They Live,}
who, after about 30 minutes into that movie,
puts on some supposedly normal pair of sunglasses,
only  to discover that ``reality'' and the ``surrounding environment,'' at least as seen through these glasses,
appears to be totally different from
what he has been conditioned to believe: wherever he looks, he reads imprinted
manipulative messages, and an alien race is intermingling with the locals.

Of the many aspects of money what is most fascinating is its ambivalence:
it can be the mediator of prosperity and welfare of individuals and  nations,
and at the same time it can be the cause of collapse of the economy.
We cannot live without money -- it is a prerequisite of our prosperity.
But at the same time it cannot be rendered without hurting us,
and presenting an existential danger to our well-being.
In what follows I shall try to argue why.

\subsection{Executive summary}

\begin{quote}
People who are experiencing inflation yearn for stable money and $\ldots$
those who are accepting the discipline and the costs of stability come to accept the risks of inflation.
It is this cycle that teaches us that nothing, not even inflation is  permanent.
({\em John Kenneth Galbraith}, in {\it {Money. {W}hence It Came, Where It Went}.})
 \index{Galbraith, John Kenneth}
\end{quote}

{{T}}{he} common citizen has been deceived into believing that money is a relatively benign, passive
and unsuspicious medium of exchange~\citep{1948-Samuelson,Begg,kyrer-penker-vwl}.
However, quite on the contrary, monetary schemes
 --  whether they are based on fiat money with no intrinsic commodity value or on commodities such as silver, copper or gold  --
present a systematic method to extract and reappropriate the wealth of nations towards institutions and goals which
would not have found public support by the common man, the middle class,
and by capitalists alike.

The gist of my argument, in a nutshell, is a pretty unspectacular claim:
Money, and in particular, ``much money,'' can be a benefactor as well as a malefactor.
Like the {\em  Lernaean Hydra} of Greek mythology,       \index{Lernaean Hydra}
or the God {\em Janus} of Roman mythology,            \index{Janus}
money has at least two aspects.
If you think you can exploit one aspect, other features show up that hurt you.
If you curb the money supply, you cripple your economy into extinction.
Conversely, if you flood an economy  with ``cheap'' money,
the greed for more and more money will devour that economy from the inside out.

On the negative side, all sorts of money  tend to get overabundant, corrupted and thus depreciated.
In the long run this excess renders money worthless
and eventually deprive people holding it.
By issuing new money and assuring new generations that this time will be different,
the cycle of emergence and corruption starts over again.
Under very general assumptions  this abuse of the power to create money is inevitable, stringent,
and thus should come as no surprise.

The main reason for the debasement  is {\em money multiplication} in various forms.      \index{money multiplication}
Examples are the degradation of coins by shrinking their physical size, or by salting them with less precious metals.
In the case of paper notes or bank accounts representing precious metals,
the authorities issue more and more notes covering less and less commodities (e.g. silver, copper and gold) per note or bit.

In the present  fiat  money system that is based on nothing but our belief in it
the amount of sound money is bound only by our imagination of its (lack of) value.
Essentially the money we are dealing with today is based on our fantasies about it alone.
Fantasies are subjective or epistemic and not objective or ontic.
As a consequence they may change easily and swiftly;
thus making (the perception of) money volatile to the follies of the mind.

What is even more disturbing is the conjecture that
{\em there does not seem to exist any alternative to money degradation.}
Any scheme suggesting to be able to curb money degradation is either unsustainable or doomed for various reasons,
all insurmountable.
That is it.
Monetary crises are an unavoidable economic destiny.
As a result, we must accept monetary crises
and  the economic tides that accompany monetary cycles.
Maybe the only dignity we have is to acknowledge this fact;
just as we have to acknowledge our own death.

Who might be the major beneficiaries of the ever increasing amounts of money?
The usual suspects are
\begin{itemize}
\item[$\bullet$]
governments and public institutions in general.
These organizations wish to receive money  without annoying their electorate either by direct taxation or by
cuts in public spending. Spending cuts and increase of taxation are always inconvenient if not outright
detrimental to the powers that be, because they represent
unpopular, painful actions which are difficult and rather tedious to implement.
The extra money created  in exchange of government debt (i.e., treasury bonds)
contributes to a general decline in value, that is, to inflation, and thus to an {\em indirect taxation;}
\item[$\bullet$]
the ``Money Trust''   --    \index{Money Trust}
consisting of all money issuing entities, in particular today's  central banks  as well as    \index{central bank}
the banking and financial industry in general, acquiring ``something for nothing''~\citep{soddy-RoM};
by collecting
\begin{itemize}
\item[$\star$]
valuable assets through monetization (see later);
\item[$\star$]
interest through the issue of fiat money in exchange of debt certificates (mostly government bonds);
\end{itemize}
\item[$\bullet$]
people holding money or assets if (and only if) they are able to realize {\em cumulative advantages} \index{cumulative advantage}
[sometimes also referred to {\em compound interest} \index{compound interest}
or the Matthews Effect~\citep{merton-68}]; \index{Matthews Effect} , as well as
\item[$\bullet$]
insiders or  lucky speculants.
\end{itemize}

On the positive side, an
abundance of money -- an economy awash in money and liquidity -- creates a ``comfort zone'' which seemingly pleases everybody.
In particular,
the public  at large  profits from an expanding economy in many obvious ways;
at least until inflation or the inevitable economic crises sets in, after which a new business cycle starts.

The situation is complicated by the fact that moneywise the economy is no zero-sum game --
sometimes all participants gain without losing; in these fair periods of high liquidity, the economy is in ``win-win mode,''
delivering wealth to everybody.
Thus it is not always true
that all sorts of money are competing for only a limited amount of goods~\citep{Wicksell-1922,Friedman-2008},
that~\citep[Chapter II]{Galbraith-money} ``other things equal, prices vary directly with the quantities of money in circulation,''      \index{Galbraith, John Kenneth}
and thus more money just means less goods per money unit; hence inflation.
For ideally more money creates more economic activity, which results in more goods,
the existence of which can motivate (via monetization, see later) even more money,
spiraling \citep{keynes-GTEIM,2006-Binswanger}  on and on.
Unfortunately, driven by (too) much  money, an economy can spiral up- as well as downwards.

The Author does not consider governments or  organizations in general as ``evil''  {\it per se,}
nor does he negate their necessity~\citep[p.~715]{von-mises_HumanActionsScholars}.
Nonetheless, he calls on the Reader to be critical about, and deeply distrust the powers that be.
Whenever they deem it in their interests   governments and all forms of institutions
react decisively, radically and brutally to extract wealth from anybody, including their own citizens.

From today's perspective it appears hardly believable  that in the last century
totalitarian dictatorships such as Soviet Bolshevism, but also
democracies such as Roosevelt's United States of America (from 1933 to 1975)
ripped off gold from, and subsequently prohibited the possession of gold by their citizens.
During the regency following Louis XIV. in France, possession of more than a small amount  (five hundred livres) of coins was forbidden \citep[p.~30]{mackey-1841}.
Just recently, the Hungarian government announced plans to confiscate
savings out of private pension funds to ``stabilize'' the roaring public deficit.
This goes on and on.
In history, sovereigns and states have stolen the wealth of their subordinates and citizens a zillion of times,
and they will do so again and again if they consider it necessary.
Often monetary policy and instruments effectively amount to more or less obvious ways to plunder the public.

\subsection{Remarks on methodology}

\begin{quote}
CAPTAIN SHOTOVER. $\ldots$ Give me deeper darkness. Money is not made in the light.\\
$\ldots$                                                                           \\
MAZZINI. $\ldots$ It's amazing how well we get along, all things considered.
({\em George Bernard Shaw}, in {\it Heartbreak House},  Act~I, \S~ii and Act~III, \S~ii.)
\end{quote}

{{M}}{oney} appears to be one of the most amazing and mind-boggling entities:
we are conditioned to its pervasion, yet we may have merely enigmatic, uncertain ideas of
how it is created, how it evolves, and even how it is being accounted.
The epistemology of money is confusing and comprises many intertwining layers of narratives;
some so trivial they resemble well-told fairy tales of deception~\citep{1948-Samuelson,Begg,kyrer-penker-vwl},
some so ``deep'' they appear to be rooted in metaphysics~\citep{soros-alchemy}.
Can we evade~\citep{Bouchaud-08} the maze or veil created by our conditioning, and erected through
(dis)information from the media, contradictory economic theories, ideologies,
and influential groups who have a vested interest in one way or another?

Indeed, \citet[Chapters I and III]{Galbraith-money} notes that  \index{Galbraith, John Kenneth}
\begin{quote}
Those who talk of money and teach us about it and make their living by it gain prestige, esteem and pecuniary return, as does a doctor or
a witch doctor, from cultivating the belief that they are in privileged association with the occult.
$\ldots$
Though professionally rewarding and personally profitable,
this too is a well-established form of fraud.
$\ldots$
The study of money, above all other fields in economics, is one in which complexity
is used to disguise truth or to evade truth, not to reveal it.

$\ldots$

The process by which banks create money is so simple the mind is repelled. With something so important, a deeper mystery seems only decent.
\end{quote}

Whatever might be the intentions of both  the Readers, the Author, and of some protagonists,
I would like to suggest to uphold an advice by {\em Sigmund Freud} \index{Freud, Sigmund} to his successors~\citep{Freud-1912},
to be aware of the dangers
caused by {``temptations to project,
what [[the analyst]] in dull self-perception recognizes as the peculiarities of his own personality,
as generally valid theory into science.''}
In this spirit, economies are treated as a client-patients, and whatever findings come up are accepted as is, without any
immediate emphasis or judgment.
It is therefore attempted not to impose any kind of ideological agenda,
besides, probably, the opportunistic
guiding principle
subsumed by the Latin phrase {\em ``cui bono?''} meaning ``to whose benefit?''    \index{cui bono}

Every effort has been made to express the following thoughts
briefly, clearly and comprehensively.
Being a theoretical physicist himself, the Author is particularly frustrated by
any vain attempt to heavily formalize and mathematize
macroeconomic processes and issues. Besides academic gratifications, such endeavors often lead to nowhere.

From a pragmatic point of view
the current macroeconomic theories have very little if any predictive power.
Indeed, maybe even a random number generator emanating advises based on pure chance would perform better!
Of course, I would not discourage formalization as such  --  just on the contrary --
but the reductionist program fails for complicated, complex and open systems which cannot be standardized and reduced
to simple behaviors  which can be composed into, or scaled to, macroeconomic scales.

Thus, very little mathematics is presented;
instead I suggest to use analytical, critical reasoning and empirical evidence as far as possible.
After all, life is too short to spoil it with an unnecessary amount of useless formalism.

I would also like to mention a {\it caveat:}
some Readers might want to wave off  the following thoughts as
sounding suspiciously like a weirdo telling you revisionist conspiracy theories.
Actually, the Author ensures his Readers that he himself has tried hard to convince himself
that ``all is well'' with our monetary system,
besides some crackpots claiming to uphold the
``true truth'' while disseminating crappy thoughts about this or that.
As matters stand, I would consider such a position to be irresponsible.

Alas, Readers who do not want to be troubled might want to
stop reading and enjoy themselves in oblivion as long as they can (afford).
If somebody wants to believe in some  {\it ``Sparefroh''} \index{Sparefroh} mentality,
I have no intention to trouble such a person.
For those not willing to, or being incapable of, providing the intelligence and willingness to cope
with certain situations, a lot of mostly unpleasant and unexpected surprises might happen throughout their lifes.
But then, they might just as well be lucky.

The apologetic title has been chosen because,
despite all shortcomings, the Author can think of no reasonable alternative to money,  in particular, to fiat money.
Thus in a very general sense he is afraid that,
despite its apparent inherent and irreducible lack of stability and the resulting inevitable crises, there is no feasible alternative to money.
And so the show goes on and on; generation after generation, and business cycle after business cycle.

\section{Money creation balanced by some thing $\quad$}

\subsection{Commodity versus fiat money}

\begin{quote}
I am time, the mighty destroyer of the world, out to destroy.
Even without your participation all the warriors standing arrayed in the opposing armies shall cease to exist.
Therefore stand up, obtain glory!
Conquer your enemies, acquire fame and enjoy a prosperous kingdom.
All these  warriors  have already been destroyed by me.
You are only an instrument.
({\em Krishna insisting upon Arjuna to fight}, in {\it Bhagavad Gita}, Chapter XI:32,33.)
\end{quote}

{{W}}{hen} it comes to ``paying'' some thing with another thing\footnote{
The asymmetry that seems to be implied by the payment is not very clear and remains purely conventional,
because why should I ``pay''
somebody for the legal acquisition of $A$ with $B$ when the other way round
-- me being ``paid'' with $A$ for the legal provision of $B$ --
appears to be another reasonable perspective?}
there are at least three possibilities:
either giving this other thing in exchange for some thing acquired,
or giving a {\em representation} of this other thing for some thing acquired,      \index{representation}
or giving something {\em believed to be generically valuable}.
Indeed, this is reflected
by the possible types of money, as
there appear to be at least three major options:
\begin{itemize}
\item[$\bullet$]
 commodity (such as gold, copper or silver) based money,
\item[$\bullet$]
 one-to-one representations in lieu of commodities,
and
\item[$\bullet$]
fiat money which has no intrinsic commodity value besides the beliefs people hold about it.
\end{itemize}

In the latter cases, the physical ``layer''
or substrate is arbitrary and not important as long as it appreciated by everybody.
Today fiat money is mostly represented by decimal digits in bank accounts,
which in turn are represented by (electric) states in semiconductors.
These states can be ``translated'' into paper money (notes) or into coins on demand at teller machines;
but mostly they are directly transformed into assets, goods, or services.

At first glance commodity based monetary schemes should be less vulnerable to inflation and degradation
because ideally the amount of money
should be strictly limited by the aggregate amount of that commodity.
Unfortunately, as history and some critical analysis shows,
in the log run commodity money appears almost as easily corruptible as intrinsically worthless fiat type money:
\begin{itemize}
\item[$\bullet$]
The commodity can be {\em substituted} by or {\em salted} with less precious substances.
In the case of gold
this might be
tungsten,  whose density (near room temperature) of
19.25~g/cm${^3}$
is almost the same as that of gold, which is 19.30~g/cm${^3}$,
but whose price is a fraction thereof.

\item[$\bullet$]
The above issues are connected to the impossibility to certify the commodity
against counterfeits in typical market situations.
How can you be sure that the grains of gold, copper or silver I present to you
for your precious something really are gold, copper or silver?
You cannot and never will be, for the capacity to certify is accompanied by the capacity to counterfeit.

\item[$\bullet$]
If, for convenience and sustainability and certification, gold or any commodity
is substituted by any kind of {\em representation} thereof -- such as notes referring to and redeemable to
gold, even in a one-to-one manner by carrying the number of an associated gold bar
--
then this substitute, which is
intrinsically worthless, is exposed to all the follies of fiat money;
in particular its multiplication by the authorities.
\end{itemize}

Despite these issues related to unfounded money multiplication,
there are other disadvantages of commodity based money:
\begin{itemize}

\item[$\bullet$]
A commodity based monetary system
is tied much stronger to the almost uncontrollable availability and abundance of that commodity
[e.g., the economically negligible production of gold from mercury through transmutation~\citep{PhysRev.60.473}]
and the resulting undesirable dependence of the amount of  money on the
aggregate amount of the commodity~\citep{Wicksell-geld-engl}.

\item[$\bullet$]
In a commodity based monetary system it is impossible to increase the money supply by the mere expectation of future profits.
As desirable this fact may be, it results in the impossibility to expand, defend and sustain the economic and
(geo)political {\em status quo.}
The resulting lack of liquidity cripples commodity money based economies with respect to others,
in particular with respect to economic expansion and military defense.
From a financial point of view, the amount of military expansion is dominated
by the arbitrary but strict limits on the commodities (mostly silver, copper and gold).
Thus eventually any such commodity money based economy will fall prey to an economy based on fiat money.

\end{itemize}

For instance, take the expansionist monetary and military policy of Nazi Germany before and during World War II.
As stated by \citet[p.~xi]{smith-gold1},
``from 1938 until 1945, $\ldots$ Hitler looted the central banks of occupied Europe.''
After the {\em Anschluss,} the German {\em Reichsbank}  \index{Anschluss}
sacked \& absorbed the Austrian central bank gold reserves,
which amounted to more than 91 tons of gold~\citep[p.~2]{smith-gold1}.
Although a catastrophe in other humanitarian and political aspects, from a purely economic point of view,
the German annexation of Austria by force was a big success from the German point of view, and a big fail from
the Austrian perspective.
All previous Austrian austerity measures directed at sustaining a noninflationary currency only
contributed to its inability to defend against an ``unfriendly takeover,''
and ultimately to the profit of its aggressor.

Thus, for pragmatic reasons, the only remaining alternative appears to be {\em fiat money} not directly backed by any commodity.
Fiat money should be intrinsically almost worthless,
making it possible to almost indefinitely expand its quantity.
The liquidity supplied to an economy by such a money volume expansion
may  result in a positive feedback loop of ever increasing production and prosperity.
However, by the same negative feedback, it may also result in (hyper-)inflation by
the restless production of additional money.
For instance, it is a mathematical fact
that the compound interest
requires excessive (actually exponential) money quantities.
In the long run, no such excessive growth of liquidity can be counterbalanced by the traded assets, goods and services.

One may argue that the supply (or increase) of fiat money should somehow be linked to the gross domestic product,
but this can be abandoned from the outright for many reasons:
there is no direct control of fiat money once the system is set ``into motion.''
Indeed, the fiat money created by the financial sector, or by the aggregate of property,
by far outnumbers any kind of economic indicator even weakly linked to the gross domestic product.
So, fiat money is only backed by the belief in and
by the fantasies people have about it alone.

In what follows, if not stated otherwise, money will be interpreted as {\em fiat money} in its various forms
-- coins and notes, states in a bank account, future promises {\it et cetera}.
Fiat money is the most convenient, flexible, dynamic and potentially explosive type of money  known today.
As we shall see, an economy based on fiat money is suspended in believe.
As beliefs change, so changes the perception and the value of fiat money.

\subsection{Types of fiat money}

 \begin{quote}
It is absurd to say that our country can issue \$30 million in bonds and not \$30 million in currency.
Both are promises to pay, but one fattens the usurers and the other helps the people.
If the currency issued by the Government was no good, then the bonds would be no good either.
It is a terrible situation when the Government, to increase the national wealth,
 must go into debt and submit to ruinous interest charges~$\ldots$.
 ({\em Thomas Alva Edison}, quoted in {\it The New York Times, December 6, 1921})
 \end{quote}

{{F}}{iat} money comes in very different forms.
In what follows two principal forms of fiat money will be discussed: fiat money created by the Money Trust,      \index{Money Trust}
and fiat money created directly by governments or other institutions, if they are authorized to do so.

\subsubsection{Historic money based on  reserve banking}

Exactly when the idea appeared to issue a {\em representation}  of some precious commodity  --   also called ``money''       \index{representation}
--
backed by some reserve of that precious commodity in the possession of the issuer of that representation,
is unclear.
It remains also unknown exactly when the idea emerged to issue {\em more} representations of precious commodities
than is ``backed by'' the  underlying ``reserve'' of that precious commodity.

The latter idea may appear both as a {\em fraud} on the receiver of this representation money,
or a {\em genial move} toward more liquidity and the resulting increase in economic activity, and thus in the generation of wealth
for everybody.
Indeed, this ambivalence towards reserve money is the basis of many fundamental debates in economics.

\citet[Chapter III]{Galbraith-money} presents                \index{Galbraith, John Kenneth}
the following fictitious history of the invention of reserve banking:
\begin{quote}
The deposits of the Bank of
Amsterdam $\ldots$ were, according to the instruction of
the owner, subject to transfer to others in settlement of accounts.
(This had long been a convenience provided by the Bank's private
precursors.) The coin on deposit served no less as money by
being in a bank and being subject to transfer by the stroke of a
primitive pen.
\end{quote}
Here Galbraith points out the convenience of  ``pen'' money bank accounts:
instead of a direct transfer of (supposedly precious) coins from the buyer to the seller,
or instead of physically
redistributing the coins at the bank
by transferring them into a different vault (supposedly by withdrawing the coins from the buyer's vault and depositing them
into the seller's vault),
the bank just reshuffles the corresponding quantities ``on paper.''
It does so by ``tagging'' the coins previously owned by the buyer with another owner, the seller,
through inscribing ``into their books'' this amount dedicated to the use of the latter.

Galbraith continues,
\begin{quote}
Inevitably it was discovered $\ldots$ that
another stroke of the pen would give a borrower from the bank,
as distinct from a creditor of the original depositor, a loan from
the original and idle deposit. It was not a detail that the bank
would have the interest on the loan so made. The original depositor
could be told that his deposit was subject to such use -- and
perhaps be paid for it. The original deposit still stood to the
credit of the original depositor. But there was now also a new
deposit from the proceeds of the loan. Both deposits could be used
to make payments, be used as money. Money had thus been created.
$\ldots$
There was that interest to
be earned. Where such reward  is waiting, men have a natural instinct for innovation.
\end{quote}
That is, now the bank -- in lieu of the original owner --
gives away ownership of the deposit to some third party, the borrower.
The motivation for this transaction is the {\em interest} paid by the borrower. \index{interest}
This interest might  somwhow be split among the original owner (for the provision of the original deposit)
and the bank (for their services in marketing and handling the loan).

Galbraith then continues,
\begin{quote}
There was an alternative opportunity involving bank notes
$\ldots$
That was to give the borrower not a deposit but a note
redeemable in the hard currency that had been placed in the
bank as capital or as a sedentary deposit.
\end{quote}
Here, finally, the bank gets rid of the original owner of the deposit and issues its loan money without
referring to the former, as long as a certain fraction of deposits is reserved for
redemption of the money notes issued.\footnote{
Statistically, the laws of probability require that
in the case of a ``large'' number of customers (in ``normal'' times of no bank runs),
the variations (variance) of the {\em aggregate holdings} set aside as reserves for instant withdrawal of these customers     \index{aggregate holding}
would only vary with the {\em square root} of the number of customers times  the variation
of an {\em individual holding}~\cite[p.~67]{Wicksell-geld-engl}.
Thus, the reserve requirements per customer decrease with an increasing number of customers.
}

Evidently the advantage of the latter scheme is that the bank does not have to share the profits from the interest payments.
And so, fractional banking might have been born.

The profit of a bank is proportional to the amount of money it can generate and issue.
Thus, because the more money can be created with the available reserves  the more
credit can be given, and the more interest contributing to bank profits can be collected.

\subsubsection{Fiat money based on fractional reserve banking as we know it}

The historic type of reserve banking mentioned above has been developed
into a supposedly more sophisticated {\em fractional reserve banking}~\citep{ModernMoneyMechanics}.  \index{fractional reserve banking}
Thereby, fiat money as we know it today
is basically generated by banks in a kind of cascade mechanism.

To governments and central bankers
the fractional reserve banking, as compared to reserve banking involving only private banks, has the advantage that, at least in principle, the
source of all money is a central bank.
In some ways central banks may thus be easier to hold accountable, control, and steer into particular {\it modi operandi} -- such as printing money
or buying treasury bonds -- than private banks.
Whether this possibility to manipulate the money supply is a desirable feature depends on the perspective, and on subjective interests.

\subsubsection*{Boot money generated by central banks}

Thereby, to boot a currency,  the primary fiat money is ``produced''  by {\em cental banks}
out of thin air.

Reserves, for instance in the form of gold or foreign currencies,
are not necessarily required for this boot process.
Nevertheless, the impression of their existence generates public trust domestically and abroad (thereby fostering trade).
Thus historically central banks were often supposed to possess ``appropriate'' amounts of reserves such as gold or other deposits of value.

Examples of central banks are
\begin{itemize}
\item[$\bullet$]
the Bank of England. It started as a private corporation that became nationalized by a Labour government in 1946;
\item[$\bullet$]
the Federal Reserve System in the United States of America      \index{central bank}        \index{Federal Reserve System}
It  is ``much like'' a private corporation.\footnote{
By its own account
postet  at URL (accessed on Dec. 16, 2010) \url{http://www.federalreserve.gov/generalinfo/faq/faqfrs.htm},
``The twelve regional Federal Reserve Banks,
which were established by Congress as the operating arms of the nation's central banking system,
are organized much like private corporations -- possibly leading to some confusion about ``ownership.''
For example, the Reserve Banks issue shares of stock to member banks.
However, owning Reserve Bank stock is quite different from owning stock in a private company.
The Reserve Banks are not operated for profit, and ownership of a certain amount of stock is,
by law, a condition of membership in the System. The stock may not be sold, traded,
or pledged as security for a loan; dividends are, by law, 6 percent per year.''
}
Its actual ownership structure remains unclear~\citep{cc-76};
\item[$\bullet$]
the Bank of Japan. It is a stock company  whose
subscribers are mainly
the government (holding 55\% of the total subscription), as well as
individuals  (holding 45\% of the total subscription).
It is also owned by
financial institutions,
public organizations,
securities companies, and
other firms; \footnote{
Taken from  page~12 of {\em The Annual Review 2010} of the {\em Bank of Japan};
URL (accessed on Jan. 1, 2011)
\url{http://www.boj.or.jp/en/type/release/teiki/ar/data/ar1002.pdf}
}
\item[$\bullet$]
the {\em Schweizerische Nationalbank} {\em (Swiss National Bank)}.
It is a stock company  currently co-owned by
2225 private and 78 public shareholders;\footnote{
Taken from p.~143
of the {\em Gesch\a"ftsbericht 2009} of the {\em Schweizerische Nationalbank (SNB)};
URL (accessed on Jan. 1, 2011)
\url{http://www.snb.ch/de/mmr/reference/shares_structure/source}
}
\item[$\bullet$]
the Eurosystem and its member banks,
such as the {\em \"Osterreichische Nationalbank} {\em (National Bank of Austria)}, which became nationalized lately.\footnote{
More precisely, the previous owners of the {\em National Bank of Austria} have been the
{\em Republic of Austria} (that  already in 2006 had acquired 11.9\%  shares from the
{\em BAWAG} and the Austrian labor union
{\em \"OGB} which held 8.7\% in a deal to rescue its bank {\em BAWAG})  holding  about 70,27\%,  
the
{\em Raiffeisen Zentralbank}  holding  about 8.73\%,
the
{\em Wirtschaftskammer \"Osterreich}   holding  about  8.33\%,
the
{\em B\&C} (previously {\em Bank Austria})    holding  about   4.27\%,
the
{\em Uniqa}  holding  about  2.67\%,
the
{\em Industriellenvereinigung}     holding  about    2\%,
the
{\em GraWe}    holding  about 0.67\%,
the
{\em Pensionsfonds der N\"O Landwirtschaftskammer}       holding  about 0.67\%,
the
{\em Niederösterreichische Versicherung}       holding  about 0.53\%,
the
{\em VIG}      holding  about 0.47\%,
the
{\em RLB N\"O-Wien}      holding  about 0.4\%,
the
{\em O\"O Wechselseitige Versicherung}   holding  about 0.33\%,
the
{\em \"Arztebank (Volksbank)}    holding  about 0.13\%,
the
{\em Kathreinbank}     holding  about 0.07\%,
the
{\em Raiffeisenverband Salzburg}       holding  about 0.07\%,
the
{\em Raiffeisen-Landesbank Steiermark}         holding  about 0.07\%,
the
{\em Raiffeisen-Landesbank Tirol}      holding  about 0.07\%,
the
{\em Raiffeisenlandesbank Vorarlberg}  holding  about 0.07\%,
the
{\em Raiffeisenlandesbank Oberöstereich}       holding  about 0.07\%,
the
{\em Raiffeisenlandesbank Burgenland}  holding  about 0.07\%,
and the
{\em Raiffeisenlandesbank Kärnten}     holding  about 0.07\%
of the shares.
}
\end{itemize}

\subsubsection*{Secondary money generated by commercial banks}

This primary fiat money is then {\em multiplied}
by private banks which -- at least ideally and in principle -- must hold a certain percentage of
the money they themselves produce (again out of thin air against some collaterals).
The {\em minimal reserve requirements}
(or {\em cash reserve ratio}) vary greatly from zero percent (0\%) fractional reserves in countries like the United Kingdom, Sweden, Canada or Mexico,
to 30\%
in Lebanon.
In the United States it is zero percent (0\%) fractional reserves up to 10\%,
depending on the size of deposits a bank holds,
in Switzerland it is 2.5\%,
in Japan it is up to 1.2\%,
and in the Eurozone  it is 2\%.

In theory, the multiplication process could go on indefinitely in a geometric progression with smaller and smaller amounts of money,
yielding a finite money multiplier which is the sum of all the money generated~\citep{ModernMoneyMechanics}.
But this is hardly realistic, given the constant tendency of financial institutions to evade the  minimal reserve requirements
by ``outsourcing'' the generated debt through {\em collateralized debt obligations} or other accounting strategies.

Commercial banks do not make much profits from ``Sparefroh'' activities such as collecting savings from their customers.
Actually, customer savings accounts might be even perceived painful, as in this case the cash flow
is negative towards that customers.
The main profits are generated by interest payments from credit creation.

\subsubsection{Government created fiat money as we do not know it}

Throughout history,
governments have always printed (fiat) money
directly, without the intermediary of a central bank
issuing fiat money in return for government (treasury) bonds as collateral.

The big difference in this regime with respect to fiat money created in a fractional reserve banking scheme as quoted earlier
is the fact that the issuing government does not have to pay any interest on the money it created itself.
Stated pointedly, money creation is not tied to the creation of public debt,
and as a result does not imply payment of interest (by the collection of taxes) to anybody, and in particular not
to central banks.
More polemically direct money creation by governments excludes {\em usury.}

Often excessive money creation
by governments,  after a subsequent inflation,
have rendered this money worthless.
The American Revolutionary War~\citep{Galbraith-money},  the French Revolution~\citep{white-1933},
and the Soviet Revolution in Russia    \index{Galbraith, John Kenneth}
are examples of political turnovers which financed the associated military and social measures with
excessively created money which subsequently became worthless; see also the quotation from
Benjamin Franklin in Chapter \ref{stgrod}.
Also non-revolutionary France after the  reign of Louis XIV.~\citep[p.~6]{mackey-1841}
printed eccessive amounts of money.

Examples for such a more stable fiat money directly created by a government (without some bank intermediary)
is  the {\em greenback} referring
to greenish paper currency that was
originally issued directly into circulation by the United States Treasury
to pay expenses incurred by the Union during the American Civil War and President Lincoln.
It was not backed by any gold standard but instead was backed by the authority of the United States government,
and indirectly by the people living and paying taxes there.

\subsubsection{Which tail wags which dog?}

Indeed, one argument in favor of the central banking way of creating money and against direct government money printing,
is the capacity of  the central bank to act, at least in principle, as just an additional power
whose independence should be assured by
the principle of {\em the separation of powers} model for the governance of states,
along the normal division of government branches into an executive, a legislature, and a judiciary.
Although breaches of this separation principle are to be expected,
they might be rare in not-so-corrupt phases of history.

However, it is not totally unreasonable that,
although this principle could be upheld for some time,
it will eventually be corrupted by so-called ``just causes.''
These comprise
\begin{itemize}
\item[$\bullet$]
preparation for  war,
\item[$\bullet$]
financing unbalanced budgets by dept as an alternative to unpopular budget cuts or tax increases, or
\item[$\bullet$]
the fight against
economic crises after speculative bubbles collapse.
\end{itemize}
All of the above contribute to money multiplication and thus to a demise of that money.

The Author takes it to be evident that no central bank, private or public, can withstand the pressure of politics in the long run.
In particular, in times of crises,
if  central banks are pressured by politics and their ``peers,''
they may (in)directly issue  ``the required amount of money'' (as seen by the government)
in exchange of government or private debt (sometimes referred to as {\em quantitative easing}).  \index{quantitative easing}

It appears almost ridiculous to pretend that (central) banks can effectively
act independently from politics.
They do not and can never be expected to do so.
For the sake of a historic evidence consider the fate of the Knights Templar.

Indeed, throughout regular times, the interplay between central banks and policy is not so visible, but
in particular in times of crises it is not difficult to realize that
central banks are at governments' and thus politicians' mercy --
at least in state owned central banks
it is an understatement that the ``political tail'' is the master to begin with.
In systems like the Eurosystem, politics goes some ways to ensure that, for instance,
at least officially the heads of the central banks are not directly linked to political parties;
but anybody with even a slight interest in these issues could be well aware of the negotiations going on between
major political and national factions when it comes to persons and strategies.
In Austria, this system is called {\it St\"andestaat} and {\it Proporz} \index{Proporz}
(from {\em appropriation} of power, privileges and resources by different political and social factions),
as opposed to the {United States of America} plutocratic~\citep{PlutonomyReport1} system of {\em lobbyism} \index{lobbyism},
and the Soviet system of privileged {\em nomenclatura.}\index{nomenclatura}

For example, the current governor of the {\em National Bank of Austria} has served as a member of parliament for the {\em Social Democrats} from 1978 until 1999.
Previous governors have been close to or associated with the {\em Austrian People's Party.}
As stated earlier, the {\em National Bank of Austria} has recently been fully nationalized;
its previous owners had also been interest groups such as trade unions, the {\em Austrian Federal Economic Chamber},
the {\em Federation of Austrian Industries} as well as some private banks and insurance companies.

Nevertheless it is not totally clear  whether, quite on the contrary to state influence on the Money Trust,       \index{Money Trust}
one might more appropriately consider some central banks as
the ``tail wagging the state.''
This may be particularly correct if
central banks are privately owned, such as the {\em Federal Reserve System} \index{Federal Reserve System}
whose past and present ownership remains unclear~\citep{cc-76}.\footnote{
The Federal Reserve System was allegedly conceived at a secretive, confidential ``duck hunting'' {\em Jekyll Island} meeting of people
related to J.~P.~Morgan, Kuhn, Loeb \& Company, the Rothschilds, the Rockefellers,
and the Warburgs~\citep{Griffin-94}.
}
There are some indications that governments are forced to accept settlements of or securities for debt,
as well as pay (compound) interest on (treasury) bonds
which are favourable for the Money Trust.              \index{Money Trust}
Nevertheless, the Author strongly doubts that -- just because government {\em defines} the {\it modus operandi} of states -- any kind of ownership structure can withstand government influence in the long run.
As noted by \cite[Chapter IV]{Galbraith-money}, ``governments keep their central banks on the shortest of leashes.    \index{Galbraith, John Kenneth}
This is true, with all others, of the Federal Reserve System in the United States,
which enjoys the liturgy but not the reality of independence.''

Whatever benign motives may there be for private central banking,
and whatever stronghold the government may exert on its central bank --
even if (in)direct manipulation can be excluded (indeed, how could it ever be?) --
the  ownership of a central bank  at least  entitles to  first-hand
(inside) information about the monetary measures taken (such as raising the interest rates).
So, the fractional reserve banking scheme, centered around a  central bank,    \index{fractional reserve banking}
might be rather profitable for those engaged in it.

\subsection{Monetization as {\em nothing for some thing}}

\begin{quote}
This is the origin of modern money as nothing
for something on the part of the legitimate user;
as something for nothing on the part of the issuer;
and as something for a promise to pay it back on
the part of the borrower, with sufficient security
to whom the issuer transferred the acquisition
of the something accruing gratis from the issue.
[{\em Frederick Soddy}, in {\it The Role of Money. {W}hat it should be, contrasted with what it has become}~\citep{soddy-RoM}.]
\end{quote}


{{S}}{uppose} somebody
organizes a society of co-operating individuals and institutions.
Obviously, any such configuration should not consist of self-sufficient monads,
but the parties should have scarce entities such as {assets,} services and products  to offer to one another;
that is,
these assets present some form of {\em value} in the mind of other agents or participants.
The recognition, negotiation and exchange of these assets take place in some {\em agora} or {\em market}.

Besides other functions of money as a {\em measure of economic values} and thus of price,  \index{measure of economic values}
a {\em unit of account},  a {\em store of value}, as well as a {\em measure of dept},     \index{measure of dept}     \index{store of value}
money is often introduced as a {\em medium of exchange} and {\em transaction}.          \index{transaction}     \index{medium of exchange}
The amount of value of an asset expressed in units of money is called {\em price.}     \index{price}

There emerge two immediate questions:
\begin{itemize}
\item[$\bullet$]
what is the value of assets, and how are the prices fixed; and
\item[$\bullet$]
how exactly did the negotiating parties obtain their money?
\end{itemize}

Let us consider the second question  first  --  that is, how do the negotiating parties obtain their money?
Quite simply, one can obtain money, say, for a bull.
That, of course, is only relegating the issue to the customer who offers this money:
from where did this customer obtain the money?
Probably by selling some hay bushels to somebody else in exchange of money, and so on and so on.
This indirect barter could go on forever without any clue about how the money was introduced into the system in the first place, provided
the economy contains enough money to allow unimpeded exchange.

Ideally, in a purely commodity based monetary system, barter could be maintained forever.
For in such a system some commodities, say gold, copper or silver, take on effectively the role of money.
Everyone in the possession of this commodity can acquire anything else,
as this commodity is accepted as barter by anybody else.
Commodity money needs no monetization, as the market value of the commodity could,
at least in ideal markets, be related to the market value of the asset exchanged,
thus fixing the price in terms of the commodity (money).

This is not the case for fiat money, because this type of medium of exchange, or (legal) tender,
has almost no intrinsic commodity value whatsoever.
So how exactly does fiat money enter the system in the first place?
The answer is through {\em monetization,}
that is, the process of converting some asset into some form of money that is generally {\em believed} and {\em accepted}
as a settlement of an exchange or a debt.
Obviously, in order to be generally accepted,
the issuing agency has to be a publicly certified and accepted {\em authority.}

Fiat money presents no intrinsic commodity value, and thus cannot be directly related by its (nonexistent) commodity value.
As a substitute of intrinsic commodity value,
some (central or noncentral) bank authority issuing the fiat money {\em guarantees} and {\em certifies} its value.
Pointedly stated, in an almost magical manner, some agency (im)prints something on a sheet of paper or digital account,
and in that manner creates money out of ``thin air.''
Henceforth, any such agency will be called {\em bank.}   \index{bank}
Examples of banks are central banks issuing central bank money (e.g., coins and bills),
private (investment) noncentral banks, or funds, creating computerized  deposit money accounts containing digits,
or  I-Owe-You's on some substratum, mostly on paper.

As stated earlier,
trust, faith \& authority is essential; else everybody would print their own money.
For various reasons also mentioned earlier, in real-time market situations
there is no reasonable way of getting rid of trust, faith \& authority.
Even gold is subject to counterfeiting; for instance by salting it with tungsten,
or even covering tungsten bars with a thin layer of gold.

For the sake of demonstration, suppose you are a cashier.
Then you surely would not take a sheet of blank paper where I just wrote ``\EUR{100}''
as down payment for a bottle of wine, returning to me some central bank notes as change;
yet you would be willing to value that same sheet of paper if it is ``backed''
by some authority  --  such as a credit card company  --  you have been conditioned to trust.

Monetization facilitates the chain of exchanges,
as banks pass on the money created to somebody possessing assets, thereby acquiring (rights on) these assets.
In the view of the asset holder, monetization is the act of {\em turning in} (rights on) assets, thereby obtaining money.
From the bank's perspective, the exchange looks conversely.
In this process, the bank acquires both the asset as well as liabilities (balanced by the ownership of the asset)
in the form of the money issued.
Note that, the bank just {\em produces} money effortlessly ``out of thin air,''
whereas the asset holder had to acquire (e.g. inherit or produce) the asset before selling it to the bank.
Surely this puts the banks in a very privileged position:
it effortlessly acquires assets and their associated utilities by producing intrinsically worthless money.
Another privilege of banks which will be discussed below is the levy of interest on debt.

After monetization, the bank can utilize the asset,
 whereas the seller and previous owner of that asset can utilize the money issued by the bank.
In one extreme case, the money may remain ``dormant''
in the bank's account, possibly without even collecting (much) interest; this would be most favorable for the bank.
In the other extreme, the seller rushes to convert the bank's money immediately into nonbank assets,
commodities (such as silver, copper, gold, or oil),
or central bank money; any such exchange would be least favorable for the bank.

Why should anyone trade in fiat money issued by banks, which is intrinsically worthless,
for an asset which has some utility?
Because under normal conditions anyone could exchange fiat money for other valuable assets and utilities.

Examples of monetization are the bank's acquisition of
\begin{itemize}
\item[$\bullet$]  government bonds (based on future government income; e.g., claims of taxes),
\item[$\bullet$]  future claims on profits of or assets in non-governmental institutions, such as corporations,
\item[$\bullet$]  real estate property,
\item[$\bullet$]  commodities,
\item[$\bullet$]  shares in a business or company,
\item[$\bullet$]  foreign money,
\item[$\bullet$]  finance derivatives, including over-the-counter (OTC) derivatives,
\item[$\bullet$]   ``Love Letters''~\citep{Sibert10,Flannery2009,Hreinsson2009}, \index{Love Letters} that is, mutual liabilities exchanged by banks
[e.g., three subsidiaries of Icelandic banks
were posting such notes as collateral at the
{\it Central Bank of Luxembourg},
receiving {\it Eurosystem} loans in exchange, on which they subsequently defaulted~\citep{ECBank2009}].
\end{itemize}

{\em Pro forma,} the insertion of money into an economy {\em via} monetization is just another exchange,
taking place between the bank and the holder of the asset without any ``intermediate'' money state;
the role of the bank's asset being played by money.
As a rule of thumb, banks will monetize everything they seem fit according to some rules
laid out by laws and regulatory bodies, and bound by ratings issued by {\em rating agencies}.    \index{rating agencies}

\subsection{Value and price of things not traded}

\begin{quote}
 Using our housing market data from the first 11 months of the year,
along with some forecasting for December, our research arm has calculated that
 U.S. homes are set to lose \$1.7 trillion in values during 2010.
$\ldots$
Since the peak of home values in June 2006, more than \$9 trillion in values has come out of the housing market.
({\em Katie Curnutte, Zillow PR Manager},
in {\it Zillow Blog ``Early 2010 Housing Stabilization Fizzles; U.S. Homes Set to Lose \$1.7 Trillion This Year''
on December 9, 2010})
\end{quote}

{{W}}{hat} does an alleged loss
of US\$~$9,000,000,000,000=$ US\$~$9\times 10^{12}$
--
or, expressed in ISO standards, 9 tera-US\$, or 9 TUS\$
--
in property values
exactly mean?
For the sake of demonstration,
suppose (very unrealistically) that none of these properties is available for sale.
Then, as these properties are not traded,
not really much happens at all.
Some home owners might get a bad feeling like ``oh my, our house lost part of its potential selling value!''
But as long as they do not speculate and use their houses for living in them (and have no mortgages on them),
the value of these houses are completely irrelevant.
This is very similar to crashes or rises of price in other markets, say stocks or commodities.
What, to take another example,
does it mean to you that your wedding ring doubled its value since the gold price doubled?

Thus, all values of goods and services which are not traded
--
and this is the vast majority thereof
--
are {\em counterfactuals} based on ``What Ifs.''  \index{counterfactual}
A {\em counterfactual} is a would-be-price or
{\em contrary-to-fact conditional}
\citep{chisholm-46}
which has not been paid but potentially could have been paid
if some market participants would have decided to do so;
alas the market participants  decided to act differently,
either by not buying or selling anything or by buying or selling a different thing.

Already scholastic philosophy,
for instance, Thomas Aquinas,
considered similar questions such as whether God has knowledge of
non-existing  things \cite[part one, question 14, article 9]{Aquinas} or things
that are not yet \cite[part one, question 14, article 13]{Aquinas};
see also Specker's \citeyearpar[p.~243]{specker-60}  reference to {\it infuturabilities}
in the context of quantum theory.

\subsection{Value and price of traded things}

\begin{quote}
Another story is told of an English traveler, which is scarcely
less ludicrous. This gentleman, an amateur botanist, happened to
see a tulip-root lying in the conservatory of a wealthy Dutchman.
Being ignorant of its quality, he took out his penknife, and peeled
off its coats, with the view of making experiments upon it. When it
was by this means reduced to half its size, he cut it into two equal
sections, making all the time many learned remarks on the singular
appearances of the unknown bulb. Suddenly, the owner pounced
upon him, and, with fury in his eyes, asked him if he knew what he
had been doing?
``Peeling a most extraordinary onion,'' replied the
philosopher.
``Hundert tausend duyvel!'' said the Duchman; ``it's
an {\em ``Admiral Van der Eyck.''}
$\ldots$
and,
notwithstanding all [[the English traveller]] could urge in extenuation, he was lodged in
prison until he found securities for the payment of this sum.
({\em Charles Mackay},
in {\it   Memoirs of  Extraordinary Popular Delusions and the Madness of Crowds. {V}olume {I}})
\end{quote}

{{F}}{or} those things traded the value or
price can be expressed and represented by the amount  of money necessary to acquire them on some market.
It is determined and fixed in a market or {\em agora,} ideally {\it via} supply and demand.
That is, the amount of money necessary to buy some thing -- its price --
is the quantitative measure  of value.

In  more practical terms, value and prices are derived from fantasies people have about  scarce assets.
Thereby it is less important what {\em really is} than what {\em people believe what really is.}
In particular in times of hype and crises,  the public perception of the scarcity or abundance of an entity is of greater relevance than the fundamentals.
In this sense, prices are {\em epistemic} rather than {\em ontologic.}

Suppose I would possess a horse, and I would develop fantasies about romantic rides in the woods;
I might get so excited and emotional about these horse riding fantasies
that my break-even point for selling this horse to somebody else (maybe with similar fantasies)
settles at a multitude of the price at which I myself  bought the horse earlier.
If I sell, I make a profit.
The exchange will go through if I can communicate, establish and realize that kind of fantasy at some market.

Recall, for example, past price rises of some inner city property,
or of some property sections close to the sea shore or to a lake.
These sections have been valued very poorly by the original farmers possessing them;
for their utilization of land was not in terms of beauty and recreation, but in terms of harvest and food.

As there are various markets with very different fantasies and utilities
--  some of them rather isolated from each other  --
many fantasies co-exist at any given time in a single economy.
The common element of the economy is the money available or created.
Since it is dependent on various asset values and prices, which itself are determined by fantasies, the amount of monetization is a dynamic, volatile quantity.
Moreover, the relative appropriation is dynamic:
it may, for instance, be possible for one group of assets  --  say, for example, stocks or other financial assets  --
 to ``overtake'' other sectors or economic segments  --  say, for example, labor salaries or property prices.
Thereby, a dynamic appropriation of money is obtained.

These mild forms of subjectivity and conventionality of value and price
are prevalent in normal times when prices vary slowly.
In times of hype and crises,
the subjective element dominates any objective considerations (such as ``utility thresholds''),
and prices surge and drop dramatically without immediate objective reason.
This behavior can possibly be best understood in terms of  mass hysteria~\citep{mackey-1841}
an positive and negative feedback loops~\citep{gloetzl-1995} leading to instabilities.
We shall come back to these issues later.

It should always be kept in mind those changes in values as well as price stability
are always accompanied by (sometimes counterfactual) reappropriation.
Any increase, stability or decrease in value takes place in a bigger market situations, in which prices
are effectively only relative measures of value.

\subsection{Limits to fiat monetization}

{{C}}{ould}
 monetization go on forever?
Essentially yes~\citep{schneider-VWLIII},
but
monetization is, at least in principle,  bound by three constraints:
\begin{itemize}
\item[$\bullet$] by the asset value, as assessed by rating agencies or otherwise;
\item[$\bullet$] by the reserve assets an economy is capable to render; and
\item[$\bullet$] by the {\em types} of necessary reserves,
as the seller may request to be paid in money (e.g.,
by central bank money, in case of noncentral banks,
or by foreign exchange  in case of central banks),
or commodities a bank is incapable to produce (such as gold, copper, silver or energy).
\end{itemize}

In particular, in {\em fractional reserve banking} schemes~\citep{ModernMoneyMechanics}, \index{fractional reserve banking}
the money creation by noncentral banks should, at least by this principle,
be bound by the inverse of the required fraction of central bank reserves~\citep{kyrer-penker-vwl}.
How much and what kind of asset qualifies as adequate {\em collateral} and {\em eligible asset} for reserves
is a matter of convention.
These conventions are listed in a somewhat lengthy and unjustified boring enumeration of all qualifying
assets; for instance the documents issued by the Eurosystem \citep[Chapter~6,7]{EUROMonPol08} for the {\em Eurosystem}, and  by Federal Reserve system
\citep{FRAC2009}.

As  experience shows
that does not prevent  central banks to deliver {\em ad hoc} money also to other entities in times of crises.
Through ``quantitative easing'' policies central banks in Japan,
Britain, the United States of America,
and in the Eurozone have bought up treasury bonds of (their respective) states,
as well as
corporate bonds and stocks.
It is one of the big mind-boggling features of the present fiat monetary system that --
most likely due to the  general trust our moneys continue to enjoy in public --
this has not (yet) led to greater inflation.


Understandably, in order to get rid of external limits,
a bank will  try to evade the restrictions originating from the reserve requirements mentioned before by various methods and techniques.
The  explosion  of the money aggregates produced by the commercial non-central banks, often denoted by M1, M2, and M3,
as compared to the coins and notes in circulation,  denoted by M0,  is a very clear indicator thereof.
In doing so, at least in principle,
any individual bank could acquire the available marketed assets by newly created money
even beyond the ``utility threshold''  --  which is bound by the interest rates  --  for private investors.
Of course, acquisition may not be a bank's primary role or concern
since after acquisition the bank would have to properly utilize the asset;
a task which it may find notoriously incapable of performing.

Note that monetization of assets depends greatly on fixing its ``intuitive'' market value in terms of a formal {\em price} (in currency units).
As assets are often not directly marketed or traded, the fixation of asset prices is consigned to {\em rating agencies}.   \index{rating agency}
If the rating agencies  are co-owned by the very banks monetizing the asset, or if they are paid by either (buyer's or seller's) side of the transaction,
a {\em conflict of interest} \index{conflict of interest} may occur~\citep{smith-walter-09}.
That is, by overrating assets systematically,
a bank my effectively be able to generate its own almost unlimited supply of money.

Note also that, as has already been shortly mentioned earlier, the ratio of money created by the central bank versus other banks can be estimated by ratios of  currency {\em components,}
serving as empirical measures of aggregates of money stock.
Bank money is often denoted by M1, M2, M3; as compared  to the amount M0 of currency, that is,
coins and notes, in circulation.
This ratio amounts to a few percent (M3 is no longer published for the U.S. Dollar), so  most of the money  is not in currency stock.
Through the {\em fractional reserve  banking} scheme utilizing the {\em reverse multiplier}~\citep{ModernMoneyMechanics},    \index{fractional reserve banking}
and through other  less benign and accountable  practices of money creation,
 --  for instance, by bundling and re-selling debt as investments which have a (triple-A or lower) rating from rating agencies indirectly belonging to the issuers  --
most of this noncash money is created by noncentral banks.

\subsection{Absence of a counterbalance between fiat money and assets}

{{I}}{deally},
the money volume should be counterbalanced by marketable assets (and future entities; see below), so that there neither is a ``lack'' nor
an ``excess'' of money, resulting in deflation and recession, or in inflation, respectively.
I leave it to the Reader to imagine how hopeless such an endeavor of the creation and maintenance of a ``balanced fiat monetary equilibrium''
in economies with dynamical creation and annihilation of assets, products, expectations, and value is.

Take, for example, the monetary side of the notorious~\citep{food-for-oil}
{\em Oil-for-Food Programm}  \index{Oil-for-Food Programm} after the First Iraq War,
and subsequent schemes pursued by the {\em Coalition Provisional Authority} \index{Coalition Provisional Authority}
after the  invasion of Iraq.
The  authorities  for selling oil could have effectively monetized oil; that is,
they could sell it to some companies or directly to banks; in exchange for oil they could
produce money ``out of thin air.''
According to some sources~\citep{360tonsofcash}
360 tons of cash in the form of US\$ notes  totaling US\$~$9,000,000,000 =$ US\$~$9 \times 10^9$
--
or, expressed in ISO standards, 9 giga-US\$, or 9 GUS\$
--
in pallets were flown from the US to Iraq in exchange for oil.
The cash was printed after BNP Paribas, on request of the United States of America, was authorized by {\em United Nations} \index{United Nations} to
transfer an equivalent amount to the {\em Federal Reserve System}. \index{Federal Reserve System}
After the oil was consumed, the cash remained in the economy and contributed to M0 (if not burnt or irretrievably lost otherwise).

From a counterbalance point of view, there exist two possibilities:
\begin{itemize}
\item[$\bullet$]
the oil could be seen as an investment for producing other assets (such as plastic) whose value might even be higher than all the assets and services consumed through its production;
\item[$\bullet$]
alternatively, the oil might just have been burnt for heating or leisure trips.
\end{itemize}
It can be safely stated that nobody knows what the ratios between these alternative is.
Indeed, this is true not only for Iraq oil, but for all oil, regardless of its origin, and for other monetized commodities in general.

Thus, whether  the monetization of assets is counterbalanced and thus contributes
to inflation or deflation remains unknown.
Yet, it is generally believed that there is a temporally stable balance between the money stock and assets.
With this economic unknowable the ``naive'' quantity theory of money characterized by the statement~\citep[Chapter II]{Galbraith-money}
``other things equal, prices vary directly with the quantities of money in circulation,''
    \index{Galbraith, John Kenneth}
breaks down, because both the ``quantities of money in circulation'' as well as the ``equality of  other things'' are concepts
which can neither be operationalized nor measured.
So, one is again relegated to the fantasies of the markets and individuals.

Indeed, the public belief in such an effective equilibrium is a necessary and inevitable delusion for the acceptance of any kind of fiat money.
As long as this delusion can be maintained for or imposed upon the majority of market participants, fiat money is relatively stable;
regardless of the disproportion between volume of fiat money and marketable assets.

\subsection{Fiat fantasies {\em versus} deficit spending}

\begin{quote}
 You do look, my son, in a mov'd sort,
as if you were dismay'd: be cheerful, sir:
our revels now are ended. These our actors,
as I foretold you, were all spirits and
are melted into air, into thin air:
and, like the baseless fabric of this vision,
the cloud-capp'd towers, the gorgeous palaces,
the solemn temples, the great globe itself,
yea, all which it inherit, shall dissolve
and, like this insubstantial pageant faded,
leave not a rack behind. We are such stuff
as dreams are made on, and our little life
is rounded with a sleep.
({\em Prospero},
in {\it William Shakespeare's The Tempest})
\end{quote}

{{I}}{n}
more recent times the money created {\it via} monetization of products of the financial sector, such as
speculative derivatives of any kind, has {\em by far outnumbered}
the money created by governments (bonds) in their attempt of deficit spending
as a means to maintain low unemployment rates, social stability, and military spending.
Therefore, it appears almost absurd when investment bankers (and some associated ``conservative'' economists),
blame governments of Keynsianism, while simultaneously
engaging in excessive fiat money creation several magnitudes higher than any of these additional government activities.
Furthermore, in times of crises they attempt to consolidate their assets by seducing or blackmailing governments
into securing them through
future taxation;
thereby effectively taking hostage entire populations, as well as the future generations,
of these states by dragging them into debt.
Indeed, besides all ideological ``blabla'' talk directed toward the uninformed public,
all kinds of money multiplications
 --  both from government sources as well as from the financial sectors  --
contribute to an inflation of the monetary base and to its effective demise.

\section{Money creation by expectations $\quad$}

\subsection{Indirect fiat monetization by expectations}

\begin{quote}
The financial industry has become a menace to society.
Its ability to create credit has brought it undue political influence,
enabling the industry to deregulate itself and to engage in such excesses
that only a massive infusion of taxpayers' money has saved it from extinction.
$\ldots$ Credit creation is too dangerous to be left to the discretion of bankers.
({\em Richard Duncan},
as quoted by {\it William Pesek}
in a {\it Bloomberg Businessweek} commentary ``Forget Geithner's Job, We'd Rather Run China'' on February 16, 2010)
\end{quote}

{{S}}{ome} nonbank agents  such as explorers, invaders, investors or inventors  require money for future profits.
Examples of such nonbank agents are homeowners expecting future salaries,
industries expecting the production of future assets,
speculators expecting a development of future markets favorable for them, or
states waging war on other states in the expectation of victory, allowing the unsolicited exploitation of the opponent's wealth.

Monetization treats the expectation of future profits quite similarly as assets:
a bank can monetize the expectation of future profits by acquiring the right of collecting paybacks from the investor in the future.
In order to make sense for the investor, these repayments should at least be counterbalanced by the expected profits.
There is a difference between a directly obtained asset and a future asset:
whereas the {\em ownership rights} of assets are immediately transferred to banks in the first, direct monetization case,
the banks obtain no immediate control over future assets.
In more concrete terms: whereas, for example, by direct monetization,
the bank can re-sell a monetized real estate property immediately after acquisition,
it could only re-sell the rights of future assets in the indirect case.
As future profits are necessarily uncertain and subject to possible failures, they are always at risk.

For a variety of reasons
 --
for instance to counterbalance their risk and the resulting unwillingness to donate money for uncertain future profits,
or to compensate for the temporal delay in consumption~\citep{Boehm-Bawerk} or foregone profits,
and as a reflection of the market price of ``generic'' future profit expectations
 --
banks levy interest.
Debt, that is, the obligation to repay in the future, is always associated with interest~\citep{pesek-saving_1968}.
Interest is the right to (regularly) collect money from the debtor, in addition to the principal  --  or to increase the principal as
the time of lending increases  --  at a certain rate.

Note that, without fiat credit and dept, the amount of money could only be sustained proportional to the growth (or decline) of marketed assets,
as at any given moment it would only be possible to invest money which has already been created,
and not also money created {\it in expectation} of future profits.
By lack of liquidity, when  compared to economies allowing  fiat credit,
this bound seriously cripples an economy.
These principles  also have to assume that through monetization (and its inverse)
a reasonable equilibrium or balance between the money stock and marketed assets can be maintained,
thereby synchronously accounting for all created and annihilated or stored assets in a sort of virtual inventory;
an assumption which is highly questionable.

Alas, if the fiat credit and thus also the debt has no direct backing
by commodities or monetized assets, the money creation is principally unbounded,
resulting in monetary crises if the future profits are overestimated.
Yet, despite these unfavorable side effects, the creation of money
through the monetization of future profits has been one of the driving forces for
the increase of production and services, and the prosperity at large~\citep{soros-alchemy,2006-Binswanger}.
Anybody arguing against monetization of future profits and fiat credit might just as well propose
going back to some kind of unrealistic and unsustainable monetary stone age.

\subsection{Interest as tax and appropriation}

\begin{quote}
A republic, if you can keep it. ({\it Benjamin Franklin} at the close of the {\em Constitutional Convention} of 1787) \\
$\cdots$    \\
The U.S., UK, and Canada are the key Plutonomies - economies powered by the wealthy.
\citep{PlutonomyReport1}
\end{quote}

{{A}}{s}
a result of usury the banking sector receives a certain amount of additional income on an annual base in terms of the interest paid.
Where exactly does this money required to pay the interest, in addition to the principal granted, come from?
It cannot come from any other source than the fiat money created by the banks themselves.

As the overall amount of valuable assets competing for money (and {\it vice versa}) is limited, the effect is a sort of general taxation by interest~\citep{champ-freeman-2004},
a re-appropriation of assets toward the banks.
Even under ideal conditions, the compound interest levied amounts to a geometric progression of the volume of money, the assets created, as well as a redistribution of wealth
in favor of the financial sector.

This dynamical reappropriation has been termed {\em {M}atthew Effect}~\citep{merton-68} because of its mentioning in the Bible.   \index{Matthew Effect}
It is commonly expressed by stating that ``the rich tend to get richer, and the poor tend to get poorer.''

\subsection{Credit as egalitarian instrument}

Contrary to the compund interest contributing to the
{\em {M}atthew Effect}~\citep{merton-68}
\index{Matthew Effect}
mentioned earlier, credit can be perceived as a big equalizer. Because,
as has been stated by \citet[p.~71]{Galbraith-money}
``It allows the man with
energy and no money to participate in the economy more or less
on a par with the man who has capital of his own. And the more
casual the conditions under which credit is granted and hence the
more impecunious those accommodated, the more egalitarian
credit is.''

Indeed, for the seller or consumer it makes no difference whether the money obtained by selling an asset,
or the product bought, has its origin in credit and debt, or in capital in the possession of the buyer of that asset,
or the seller of that product.

That is, there are really two aspects of money creation through credit and debt:
on the one hand,  money creation through credit and debt tends to tear society apart by making the poor even poorer, and the rich even richer.
On the other hand  money creation through credit and debt allows the poor to participate, ascend and come out of poorness by
invoking money which they originally do not possess.

\subsection{Consequences of no or low interest}

\begin{quote}
Thus far we have dealt with actual substances; but some forms of wealth deceive our eyes and minds alike.
I see there letters of credit, promissory notes, and bonds, empty phantoms of property, ghosts of sick Avarice,
with which she deceives our minds, which delight in unreal fancies;
for what are these things, and what are interest, and account books, and usury,
except the names of unnatural developments of human covetousness?
$\ldots$
What are your documents, your sale of time, your blood-sucking twelve per cent.  interest?
These are evils which we owe to our own will,
which flow merely from our perverted habit,
having nothing about them which can be seen or handled, mere dreams of empty avarice.
Wretched is he who can take pleasure in the size of the audit book of his estate,
in great tracts of land cultivated by slaves in chains,
in huge flocks and herds which require provinces and kingdoms for their pasture ground,
in a household of servants, more in number than some of the most warlike nations,
or in a private house whose extent surpasses that of a large city!
After he has carefully reviewed all his wealth, in what it is invested,
and on what it is spent, and has rendered himself proud by the thoughts of it,
let him compare what he has with what he wants: he becomes a poor man at once.
({\it Lucius Annaeus Seneca} in {\em On Benefits}, Chapter VII, Section X.)
\end{quote}

{{I}}{n}
view of the possible imbalances from the accumulation of wealth by the financial sector, attempts have been made to abandon interest altogether.
Christianity has condemned
{\it  usury }~\citep{noonan-57,Weiss-Zinswucher-Christ}. \index{usury}
Also Islamic communities reprehend  {\it  riba}~\citep{Hanke-Zinswucher-Islam}. \index{riba}

Alas, an abandonment of interest implies at least two undesirable alternatives:
\begin{itemize}
\item[$\bullet$] either the amount of credit has to be limited    by   criteria  which effectively introduce privileges:
if there is a limited supply of credit, who should receive it?
\item[$\bullet$] else if there is no limit to the amount of credit available, any agent in the market would find it possible, at least in the extreme case,
 to ``buy up all available assets.'' Because of the zero cost of borrowing this could happen without penalty.
But then, if there are more than one agents competing in the market, prices will go up {\it ad infinitum;} effectively causing (hyper)inflation.
\end{itemize}

For example, the high demand for real estate properties reflects the particular importance and the relevance  of housing to individuals and families.
From the point of view of the buyer, the price of a property appears to be limited by the portion of the household income available
for the payment of dept accepted for acquiring that property.
In more concrete terms, the product of  interest rate times the property  price should not exceed the
buyer's available income, and thus the price of the property is bound by  the income divided by the interest rate.
As a result, property prices tend to increase on decreasing interest rates,
as potential buyers can afford to bid higher prices.
The leverage or ratio of this price increase is determined by the inverse interest rate.
In the (absurd) limit, with ``free credit'' associated with zero interest rate,
a single buyer would be able to bid an unlimited price for any given property.
As stated earlier, by unrealistically assuming those prices will not go up due to competing money,
the buyer could acquire all properties available on the market.

\subsection{Plasticity of interest rates}

{{W}}{hy} should a noncentral bank not offer loans
with  lower interest rates than a central bank?
Surely, if a noncentral bank is bound by some reserve constraints,
it may not be totally unreasonable to offer much lower interest rates which are bound by that reserve constraints (in percent of the minimum reserve fraction).
For a bank can create money for loans virtually without any costs ``out of thin air.''
It may use this ability to compete on the loan market.

Related to these issues
appears to be a common belief that it is possible to  curb the money supply by regulatory measures.
Indeed, interest rates of consumer  credits or mortgages and, say, the U.S. {\it federal funds rate} might be correlated.
As has been mentioned earlier, this is usually explained by money volume constraints on the noncentral banks, effectively established {\em via}
some regulatory mechanisms involving bank reserves:
in order to prevent bank runs or an unbounded lending policy,
banks usually should not be able to create more money than a certain percentage or fraction of some securities or reserves they hold;
resulting in a mild form of reserve multiplication~\citep{1948-Samuelson,Begg,ModernMoneyMechanics}.
Alas, in view of the recent events connected to the packaging and reselling of dept by the financial industry in the U.S. and elsewhere,
this {\em fractional reserve banking system,}  \index{fractional reserve banking}
appears to be a commonly told fairy tale.

On the contrary, it is in the legitimate interest of banks to avoid any reserve constraints, by any quasi-legal possibility.
In the present competitive and highly liquid financial market environment,
it is impossible for financial institutions to avoid stretching the regulatory bonds to the extreme;
otherwise they will be out of business soon, overtaken by the competitors which attract their greedy investors.

As has also been mentioned earlier, the amount of outstanding credit of a financial institution is directly proportional to the interest it levies, and consequently to its income.
There is, for instance, no immediate reason why a bank should not create money and lend it out for a lower interest rate than the central bank,
provided it is not too much bound by minimal fractional reserves:
even if the interest rate is arbitrary low, as long as it is positive, there is some obtainable gain.
Likewise, no customer needs to fail because of defaulting credit:
in the extreme case it would even be conceivable to levy no interest at all until such time when the customer can serve the interest again.
Indeed, the customer may be released from debt totally and permanently; this, however, should
be done  in secrecy, because otherwise all debtors would attempt to default as well.
Such gifts, of course, can only be granted because the cost of money creation for banks is negligible.

One may argue that the levels of interest, as well as the ``haircuts'' of customer credits discussed here are inacceptable.
This may be right, but not because of the money volumes thus reappropriated to the failing bank clients.
Indeed, as the value of fiat money is primarily
fantasy-driven,
any such measure would have demoralizing effects on the perception of money,
and may thus contribute to the demise of it.

\section{Instabilities \& Unknowables $\quad$}

\subsection{Is interest sustainable?}

\begin{quote}
And I sincerely believe $\ldots$ that banking establishments are more
dangerous than standing armies; and that the principle of spending money
to be paid by posterity, under the name of funding, is but swindling
futurity on a large scale.
({\em Thomas Jefferson}, from a {\it Letter to John Taylor, dated May 28th, 1816,}
in {\em Memoirs, Correspondence and Private Papers of Thomas Jefferson, Volume IV.})
 \end{quote}

{{T}}{he}   (exponentially)   fast  growing {\em compound interest} \index{compound interest}
suggests that, because
of the resulting exploding money volume, money creation by debt and successive debt restructuring appears to be unsustainable.
Indeed, one could suspect that (compound) interest is a malversation both of governments
-- who profit from not having to levy taxes or imposing budget cuts --
and of the Money Trust, collecting that (compound) interest.
Thus it is quite justified to ask whether there are benign scenarios involving (compound) interest?

One way to avoid this vicious circle is sustainable growth:
in this scenario
the investment made possible by debt (at least) pays off
the principal {\em as well as} the (compound) interest at later times.
Thereby, the newly created assets and services can be monetized, and  (part of) the money created through direct monetization is
then used to pay back the principal debt and the (compound) interest.
Ideally, in this scenario, debt and the associated interest can create growth and wealth~\citep{2006-Binswanger}.

From this point of view, an economic expansion is only
sustainable if it is the result of an increase in investment that, {\em in addition to} an increase in saving,  is funded
by a credit expansion which creates marketable assets which at least compensate the money created.
As a result, an economic boom could in principle be sustainable even if it is only the result of a credit expansion.
If, on the other hand, the investment does not yield enough to pay back the principal and the (compound)
interest, then the money created through debt will contribute to an unsustainable Ponzi-type pyramid game, resulting
in more and more debt, and ultimately in a business cycle crash.

This position, which allows for money and credit creation,
is in contradiction to the position
 --  sometimes (misleadingly) associated with the mainstream economist's understanding of {\em Austrian business cycle theory}~\citep{Tempelman-2010} \index{Austrian business cycle theory}  --
that an economic expansion is only
sustainable if it is the result of an increase in investment that is entirely funded by an increase in savings alone.
As a corollary, an economic boom that would merely be the result of credit expansion should not be sustainable.

Nevertheless, for various rather obvious reasons also discussed earlier, it is almost impossible to adjust and fine-tune the money expansion via credit creation
with the assets to be created in the future.
During the early stages of any Ponzi-type pyramid games, expansion of credit will increase the profits of the money creating entities,
thereby effectively discouraging more prudent strategies.
Together with the pressure from governments to allow  greater fiscal deficits at  low interest rates,
the money issuing authorities will thus most likely end up with an unsustainable level of credit expansion.
As a result, these economies will eventually undergo a monetary crises as a direct consequence of unsustainable credit expansion.

\subsection{Containment and osmosis of money types}

{{I}}{f}   the markets are relatively isolated,
the (re)appropriations
from the different price growth in different sectors (e.g., food and real estate)
may not be perceivable for some time~\citep[p.~396]{von-mises_HumanActionsScholars}.
For instance, ``Gordon Gekko'' a financial {\em Wall Street} tycoon,
earning ``a lot of money'' through his financial transactions,
will not influence the prices of sausages sold on  {\em Wall Street}
``too much,'' as he might not be interested in buying ``too many'' sausages there,
simply because he cannot eat ``too many'' of them to influence the ``street price'' market for sausages. (He seems to prefer steaks anyway.)

As historic examples show, it
initially takes some time to make people realize that the money they hold looses more and more purchasing power.
(If hyperinflation is in full swing, people are much more aware of the situation.)
Indeed, the stronger stratified a society, the less will fantasies in one sector  leak through
and affect prices in other sectors.
Nevertheless, in the long run, the different market segments or sectors tend to connect through the monetary base.
Thus eventually the fantasies exerted in one of them will  diffuse  into other sectors almost like  osmosis
through small interconnections~\citep[Section~1(f)]{Friedman-2008}.
If, for instance, the same {\em Wall Street} tycoon attempts to  take over  most sausage stands of {\em Manhatten,}
the very high price he may have to pay for them may indirectly (through the rate of return on investment) affect the street price for sausages there.
In reaction, as inflation (in terms of sausage price)  goes up, labor costs will increase, contributing to a spiral of inflation.

Effectively, the system  of interlinked assets, products and services
--  starting from  precious commodities  such as gold, copper, silver, or oil, and continuing with the various money components with increasing ``virtuality''  --
resembles the multitude of  rubles  in the late Soviet Union; there the  gold ruble  would hypothetically buy 0.987412 grams of gold, yet was never available to the general public.
Also today,  one of the most harmful  market transactions would be the {\em direct exchange} of the  most abstract  form of money,
such as, for instance, government bonds or financial derivatives, into gold, copper, silver or platinum.

At the time of writing, once again~\citep{ReinhartRogoff}
the monetary system is inflated (flooded) by monetized fantasies created primarily by the financial institutions,
and to a lesser extent by all sorts of government spending, in particular wars~\citep{BonnerWiggin,StiglitzBilmes}.
In order to back up the huge unsustainable debts, the financial institutions have
turned for government and central bank help~\citep{dunken-crises2}.
Governments effectively back derivatives and other monetized fantasies created by the financial institutions
by (future) tax revenues;
whereas central banks everywhere seem to monetize government or even  private debt.

It is not totally unreasonable to speculate that the
magnitude of these transactions can neither be sustained
nor contained to the market sectors in which they have
been created. Thus the massive amounts of money
volumes (in M3) will eventually spill over to the consumer sectors,
thereby causing (hyper)inflation
everywhere.

Moreover, the capacity to
back up the M3 volume by taxation and inflation will decrease for socio-political reasons.
The electorates and societies will grow weary of saving financial institutions through acquiring their risks and debt.
Some or all ultimate debtors such as big governments, their associated
central banks, and eventually also the commercial banks will then be forced to submit to
strategies to get rid of debt.
As outlined in Chapter~\ref{stgrod} this will mostly be accomplished by inflation and by defaulting dept.

There may be several ways to survive this monetary meltdown; some of them rely on gold
(not on gold options on paper, but on self-storing the precious metal) or other commodities, some on real estate;
some people even rely on
little bottles of alcohol and strong spirits, for which, as they claim, demand will be high in times of crises.
In principle, one could make huge profits by going into debt now,  buying commodities and real estate.
After the monetary meltdown and the following consolidation of the currency,
presently acquired debt could then be easily repaid in terms of the new money.
If that sounds strange, consider buying a real estate property today
for its price thirty years ago.

\subsection{Pricing ambiguities}

{{T}}{he}  belief
that the equilibrium between supply \& demand will in general settle at a single particular price,
and the idea that there exist equilibriums in economies in general,
is an idealistic illusion.
Let us just consider a few reasons why.

\begin{itemize}
\item[$\bullet$]
The modern markets are driven by whatever {\em communication} and {\em (dis)information} is fed into them:
\begin{itemize}
\item[$\star$]
Those who control the media (i.e., those possessing and paying through ads) might attempt to control the market and public policy.
\item[$\star$]
The market participants might suffer from an overload of information;
indeed, the excessive push of large amounts of raw data on to individuals and institutions
might result in the false impression of well-informedness while actually contributing to greater confusion.
\item[$\star$]
The market participants might lack reliable criteria or authorities to evaluate the information presented to them, or might be fed with disinformation.
\item[$\star$]
The perpetual flow of spontaneous news and opinions via the media may make impossible the formation of a ``communication equilibrium.''
\end{itemize}
\item[$\bullet$]
Markets may be subject to  trade policies and military deployment which might enforce  prices.
\item[$\bullet$]
The intra-market dynamics might not be sufficiently efficient to settle prices; or there may be no convergence toward a {\em single} price,
but rather price cycles and other more chaotic regimes.
\item[$\bullet$]
The volume creation and annihilation of money and debt by governments, (central) banks,
corporations and individuals might not allow a stabile settlement of prices by creating (expectations of) a chaotic, or alternating, or ``spiraling'' regime~\citep{Kauffman198753}.
\item[$\bullet$]
As money and its various forms and derivatives
is itself marketed, the price of money becomes recursive, self-referential and reflexive.
This will be discussed in the next section.
\end{itemize}

\subsection{Formal incompleteness of macroeconomics}

\begin{quote}
It requires some intelligence to acknowledge one's own dumbness.
({\em Heimito von Doderer}, in {\it Die Strudelhofstiege}.)
\end{quote}

{{R}}{elated}
to the the impossibility of equilibria is the fact that no complete macroeconomic theory
(or ``maximal trading strategy'') exists.\footnote{
Readers not familiar with formal incompleteness might find this argument strenuous. They might consider to skip reading the entire chapter.}
In what follows it is proved by reduction to the halting problem~\citep{godel1,turing-36,davis-58,smullyan-92}
that every macroeconomic theory strong enough to contain {\em substitution,}
{\em self-reference} and (Peano) arithmetic is {\em incomplete} in the sense there exist provable {\em true} sentences which are {\em not derivable}
and thus {\em independent} of that macroeconomic theory.

The scheme of the proof by contradiction is as follows:
the existence of a hypothetical macroeconomic prediction model (trading strategy)
capable of solving the problem of whether or not a certain halting state in macroeconomics (e.g., a certain price) will occur,
is {\em assumed.}
This could, for instance, be a winning tactics of some suspicious super-trading strategy
which takes the code of an arbitrary macroeconomic theory as input and outputs yes or no,
depending on whether or not the macroeconomic theory predicts a particular halting state.
One may also think of it as a sort of oracle or black box taking in an arbitrary
trading strategy in terms of its symbolic code, and outputting one of two symbolic states, say yes or no,
referring to reaching or not reaching its desired goal, respectively.

Based on this {\em hypothetical macroeconomic prediction model (trading strategy)} we construct another {\em diagonalization (trading) strategy} as follows:
upon receiving some arbitrary {\em macroeconomic theory (trading strategy)} code as input, the {\em diagonalization (trading) strategy}
consults the {\em hypothetical macroeconomic prediction model (trading strategy)} whether or not this
{\em macroeconomic theory (trading strategy)} reaches a macroeconomic halting state (e.g., a certain price); and upon receiving some answer it just does the {\em opposite:}
if  the {\em hypothetical macroeconomic prediction model (trading strategy)} decides that the {\em macroeconomic theory (trading strategy)}
{\em reaches a particular macroeconomic state (e.g., price),}
the {\em diagonalization (trading) strategy}  {\em counteracts} that prediction (it may do so easily by some kind of market intervention, such as volume leveraged buying or selling).
Alternatively, if  the {\em hypothetical macroeconomic prediction model (trading strategy)} decides that the {\em macroeconomic theory (trading strategy)} does {\em not reach a macroeconomic halting state in a (e.g., a certain price),}
the {\em diagonalization (trading) strategy} {\em steers the economy into that state}.

The {\em diagonalization (trading) strategy} can be forced to execute a paradoxical task by
receiving {\em its own program code} as {\em macroeconomic theory (trading strategy).}
Because by considering the {\em diagonalization (trading) strategy,}
the {\em hypothetical macroeconomic prediction model (trading strategy)} steers the {\em diagonalization (trading) strategy} into
{\em  a  macroeconomic halting state (e.g., a certain price)} if it discovers that it {\em does not halt;}
conversely,  the {\em hypothetical macroeconomic prediction model (trading strategy)} steers the {\em diagonalization (trading) strategy} into
{\em  a state different from the halting state} if it discovers that it {\em reaches a halting state.}

The contradiction obtained in applying the {\em   diagonalization (trading) strategy} to its own code proves that this program,
and in particular the {\em hypothetical macroeconomic prediction model (trading strategy),} cannot exist.

\subsection{Conjecture of balanced trading}

{{I}}{t}
may be conjectured that all trading strategies are at most balanced when confronted with all market situations.
Stated differently, a trading strategy which would be able to win more
often than loosing if the market situation is varied cannot exist.
Of course, that would not exclude certain winning strategies for specified market situations.
For instance, in a bull or bear market trend resulting in  increasing or decreasing prices it is favorable to bet on  increasing or decreasing prices, respectively.
But any such strategy would fail miserably if the market trend reverses.

\subsection{Fiat money games}

\begin{quote}
Most dangerous of all would be democracy. The Bank of England was the instrument of the ruling class.
Among the powers the Bank derived from that ruling class was that of inflicting hardship.
It could lower prices and wages, increase unemployment.
({\em John Kenneth Galbraith}, in {\it Money: Whence it came, where it went}, Chapter IV.)      \index{Galbraith, John Kenneth}
\end{quote}

{{F}}{iat} money is unbounded by any supply constraints, as it can be produced almost cost-free.
Thus, as long as the  market participants can be manipulated to believe in and trust the authorities issuing fiat money
 --  and what other choice do they have when the possession of gold is prohibited  and the use of silver and copper for money has been abandoned?  --
its volume can be expanded and contracted at will of those in control, at least as long as command over the money volume is possible.

It is not too unreasonable to speculate that those who control the money supply
need not always and necessarily serve the public good
but rather in their own interest.
Alas, even if uniform benignancy is assumed, what is good for one group may not serve the interest of other groups;
hence there always will be ambivalent tradeoffs and
dynamical reappropriations.
In what follows a few schemes are enumerated which could be used for the reapproriation of the wealth of nations and within  economies.

One may argue that commodity based money did not prevent business cycles either,
and that overheating an economy by unsustainable credit creation is not limited to fiat currencies only.
Nevertheless, it might be argued that fiat currencies are more vulnerable to economic crises,
and bubbles are both greater and longer; and a subsequent correction tends to be more painful than in the commodity based case.
Alas, it may be countered that this is the price for liquidity and prosperity that comes with fiat money.

\subsubsection{Business cycles of inflation and deflation}

By alternating periods of low and high interest rates it is possible to extract assets from an economy:
{\em inflationary periods} are characterized by low interest rates,
which encourage ``bubble buildups'' through investments with very low yield at high (with respect to subsequent periods of high interest) prices, as credit money is cheap.
Subsequent {\em deflationary periods} can be initiated by rising interest rates, which cause corrections (lowering) in prices, credit defaults, and a burst of the investment bubble created
in the previous inflationary period.

By subjecting a market to successive periods of inflation and deflation, it is possible to acquire great wealth by inside knowledge:
an inside investor would be prudent to invest into assets (e.g., stocks, derivatives and real estate)
shortly before or at the initiation of inflationary periods,
riding high on the wave of increasing profits and prices,
and, shortly before a deflationary period sets in,
pulling out of these investments, thereby converting them into save havens like
commodities (e.g.  gold, copper and silver); remaining in these conservative positions until just before a new inflationary period begins.
In that way, by controlling the business cycle through money supply, more and more economic assets can be acquired almost risk-free.

This economic cycle strategy almost appears as if it represented a {\em perpetuum mobile;} but it is not, as it can only succeed if the investor is capable of controlling the money supply.
Alas, central banks, and the fractional reserve banking system through multiplication by fractional reserve banking  as a whole,  \index{fractional reserve banking}
have been institutionalized  and should be capable of doing just that
 --  controlling the money supply  --  at least
if they are independent and not constraint by politics and external monetary constraints.

One could speculate that a proper price for investment money (i.e., interest for credit) might be definable
by the {\em absence} of business cycles of the type described above.
Whether or not such an ideal interest rate exists remains highly questionable, as value and price appear highly subjective entities in particular also
if they are determined by future expectations.
Monopolistic central banks, and the fractional reserve banking system as a whole,    \index{fractional reserve banking}
merely represent a particular way of handling these issues.

\subsubsection{The perpetual money machine}

As expressed by Franklin (see next chapter) and detailed by \cite{Galbraith-money},      \index{Galbraith, John Kenneth}
the unbounded creation of fiat money can be a perfect vehicle to finance revolutions and wars of any kind.
The American, the French and the Bolshevik revolutions all used money to execute its goals.
This moneys soon became worthless, but not fast enough to crash the revolution.
After they had taken power the new regimes always tried to stabilize their money through the introduction of a new
monetary regime, thereby abandoning the old one.
People tend to forget after about a generation that this     ever happened, and the game can start over again.

\subsection{Strategies to get rid of debt}
\label{stgrod}

\begin{quote}
And indeed the whole is a Mystery
even to the Politicians, how we have been able to continue
a War four years without Money, and how we could pay
with Paper, that had no previously fix'd Fund appropriated
specifically to redeem it.
This Currency, as we manage
it, is a wonderful Machine. It performs its Office when
we issue it; it pays and clothes Troops, and provides Victuals and Ammunition; and when we are obliged to issue
a Quantity excessive, it pays itself off by Depreciation.
({\em Benjamin Franklin }, from a {\it Letter to Samuel Cooper, dated April 22nd, 1779)}
in {\em The writings of Benjamin Franklin. {V}olume {VII}.}))
\end{quote}

{{N}}{ot} everybody can get rid of debt by eliminating his creditors by killing them.
Historically, this style was executed, for instance,   in order to restore the depleted
{\em  Roman Treasury (Aerarium)}
through the Roman
proscription -- the persecution of the rich -- enacted by {\it Sulla} in 82~BC,
followed by {\em Catilina} in 63 BC who also promised  the abolition the universal cancellation  of debts inscribed on the {\it novae tabulae},
and another proscription under the {\it Second Triumvirate}  in 43~BC.
A different historic example is the fight of Philipp IV of France against the Knights Templar in 1307.

In what follows, three ``more civilized'' modes of getting rid of debt  are mentioned:
the first option is direct
monetization of debt accompanied by (hyper)inflation.
The second, politically difficult, option attempts to
cut back on spending, such as levying higher taxes, or by direct budget cuts.
Finally, the ultimate fallback option is
bankruptcy or ``haircuts'' by paying less back than borrowed.
Unfortunately, these seem to be the only long-term options, as there is no free lunch.

\subsubsection{Monetization of debt and (hyper)inflation}

{\em Inflation,} and even more so {\em hyperinflation,} is one of the major processes to get rid of debt.
The basic idea is quite simple: if debtors are able to keep interest payments at sustainable levels
in the first time of the loan,
they need not bother about the principal and the interest payments
at later times: because, relative to future price and income levels, inflation will ``melt away'' both the principals and interest;
that is, quite literally,  debtors could pay back the principal and interest from their ``pocket money.''

That kind of strategy is employed at all scales -- at least subconsciously it is adopted by small investors
acquiring home loans, and up to the government level.
In general, the higher the inflation, the faster is the relative reduction of debt,
as measured in absolute debt
divided by the absolute income;
but also the more difficult it is to sustain payments of interest in the first time of the loan.

\subsubsection{Cutbacks in spending}

Another possibility to get rid of debt is by allocating resources (such as taxes) to pay off debt.
Despite the fact that these measures are politically difficult to impose,
they are also bad for business activities, as this decrease of liquidity reduces the amount of money available in the economy
and thus might contribute to recession.

\subsubsection{Bankruptcy and ``haircuts''}

A third possibility is to write off debt by bankruptcy or partial remission of a debt, called ``haircut'' nowadays.  Here the difficulty lies on the borrower's side;
if borrowers are large banks, this might have negative influences on all kinds of business activities,
and may again cause recession.

\subsection{Anything that can go wrong, will go wrong}

 \begin{quote}
I have not failed. I've just found 10,000 ways that won't work.
 (quote attributed to {\em Thomas Alva Edison})
 \end{quote}

{{T}}{he}  various forms of money multiplication
eventually degrade and annihilate the money supply.
I will not bore the reader with historic evidence;
there exist excellent introductions into the subject~\citep{Galbraith-money,davies-hom}.           \index{Galbraith, John Kenneth}
[See
also ~\cite[Chapter~XXXI]{von-mises_HumanActionsScholars}' account,
as well as~\citet{Zarlenga-hom}'s and~\citet{Deutsch-2006}'s German  controversial reviews.]
Let me just enumerate several ways of wacking the monetary system;
every reader can point out other petty schemes.

\begin{itemize}
\item[$\bullet$]
If money is physically presented by commodities considered precious and very valuable, such as gold, copper or silver coins,
debasement is achieved by putting  less and less  of that commodities into a unit of money while maintaining its nominal value.

\item[$\bullet$]
If money is based on commodities considered precious and very valuable,
which are {\it indirectly represented} by notes {\it directly referring} to the commodities
with an allegedly fixed amount of commodity per note
 --  possibly associated with the promise to
redeem the commodity upon presentation of the note  --
degradation is achieved by printing more and more notes referring to a constant
amount of commodities.
This is often referred to as {\em commodity based money.} Finally,       \index{commodity based money}
\begin{itemize}
\item[$\star$]
seduce the public into not using, and not hoarding the commodity by making them believe it is worthless;

\item[$\star$]
stop redeeming the notes for the commodity;

\item[$\star$]
prohibit the private possession and use of that commodity.

\end{itemize}

\item[$\bullet$]
If money is partly based on commodities considered ``precious'' and ``very valuable,''
do so as before: print more notes per amount commodity.

\item[$\bullet$]
If fiat money  is based on the pure believe in it, print more money by multiplying it in various ``intelligent'' ways;
e.g.,
\begin{itemize}

\item[$\star$]
by the {\em fractional reserve banking} scheme~\citep{ModernMoneyMechanics} \index{fractional reserve banking};

\item[$\star$]
by writing {\em Love Letters} \index{Love Letters}  --  that is, by mutual exchange of ``I Owe You''s  \index{I Owe You}
and using these Love Letters as collaterals for money via monetization
if you can find a rating agency \index{rating agency}
and a central bank \index{central bank}
willing to cooperate~\citep{Sibert10,Flannery2009,Hreinsson2009};

\item[$\star$]
by packaging (e.g. household or mortgage) debt into bonds, if you can find a rating agency \index{rating agency} willing to cooperate,
and a customer willing to buy this bundled debt.

\end{itemize}

\item[$\bullet$]

by counterfeits; indeed the only difference between counterfeit and ``legal''
multiplication appears to be the source of money.
But even in this respect the differences tend to become obscure, as for instance during World War II
allegedly the {\it Banca d'Italia} \index{Banca d'Italia} counterfeited its own money by issuing the same securities
twice with identical registered numbers and codes.

\end{itemize}

There are numerous other ways to whack the money supply; the aforementioned are just commonly used examples.
Many of the world's most clever minds  --  in particular also physicists and mathematicians allured by relatively
large salaries and benefits  --  breed over new schemes to legally multiply the monetary base, thereby
contributing to its deterioration of the monetary base.
Microeconomically this makes sense; after all if you don't do it,
somebody else surely would.

Money multiplication effectively yields a general taxation of the public  in favor of the issuer of money.
The reason for this is that the former can buy less with the amount of money they hold,
as the overall price of assets and goods eventually increases.

Money multiplication also effectively amounts to a pyramid (Ponzi) scheme,
as the earlier money buys more that the later money, due to an increase of it.

The only theoretical but impractical way of avoiding these pitfalls is to stick to the physical commodities as a medium of exchange.
Alas, for a variety of reasons, e.g., for large-scale international trade, for security reasons,
as well as for counterfeiting and authentification reasons
(``who knows if this really is gold and not counterfeit?'')
this is hardly impossible in a general setup.
For instance, who assures that ``electronic gold''
is really entirely backed by physical gold, and not just multiplied and thus merely
partially covered?
Whom could you trust? It is not too unreasonable to suspect that, in
the long run, nobody and no institution  --  public or private  --  can be trusted.
Government authorities, in particular, have a very bad record in  money multiplication.


\section{Afterthoughts $\quad$}

\subsection{Desiderata}
\begin{quote}
It is not enough to have no concept,
one must also be capable
of expressing it.\footnote{Es gen{\"{u}}gt nicht, keinen Gedanken zu haben:
man mu\ss ~ ihn auch ausdr{\"{u}}cken k{\"{o}}nnen.}
({\em Karl Kraus}, in {\it Die~Fackel~697,~60~(1925)})
\end{quote}

{{W}}{hat} kind of appropriation of wealth should we adopt?
Should we, for instance, employ a ``Robin Hood strategy'' by confiscating from the rich~\citep[Chapter XXXII]{von-mises_HumanActionsScholars} and giving the poor?
Or should we adopt a strategy to promote achievement?
We should take it for granted that, regardless of the honorable motives which were present originally, all types of strategies will ultimately get corrupted in one way or another.

But even disregarding the obvious abuse of various forms of (re)appropriation,
it is the Author's conviction that, due to a lack of absolute criteria,
there is no objective answer to the question how the wealth of nations should be appropriated.
Thus all attempts in one way or another must necessarily and inevitably remain subjective.
The following two presentations are rendered for the sake of corroboration of this thesis.

\subsubsection{Heritage}
In a 1981 meeting on ``The Worldwide Consequences of Nuclear War'' in Sicily,
Paul Dirac noted that the ``capitalistic'' versus the ``communistic'' forms of economies should not wage war against
one another by maintaining two seemingly contradicting positions:
\begin{itemize}
\item[$\bullet$]
than every child should have an equal right to pursue happiness and well-being; independent of his or her origins;
\item[$\bullet$]
that the parent should be given the right to pass on to their children those privileges and wealth which they were able to accumulate;
so that their children would benefit from their parent's achievements.
\end{itemize}
Both positions may be quite justifiably considered true; yet they contradict each other entirely.
The Author cannot offer any reasonable solution to this conundrum.
I believe that it will remain with us forever.

\subsubsection{Optimizing the distribution of happiness and well-being}
One may, for instance, vainly indulge in sophisticated schemes of
quantifying and ``measuring happiness'' or of ``subjective life satisfaction'' in an attempt to
``(re)appropriate'' happiness or life satisfaction.
By some subjective scheme or belief system one may then, guided by these principles,
somehow arbitrarily assume some ``cumulative fitness function'' as a criterion for a desirable state of economic and political affairs.

Any such scheme is based on assumptions about the specific dependency of happiness on income and wealth.
The degree of happiness as a function of income has been suggested to be logarithmic;
alternatively, it has been claimed that happiness
grows linearly with income and plateaus at a certain income level.

All of these schemes and the resulting political and economic strategies lack objectivity and rational rigor.
This, again, leads to the conclusion that interventionalism \index{interventionalism}
appears inappropriate~\citep[Part Six]{von-mises_HumanActionsScholars}.
On the other extreme of economical tactics is {\em laissez faire}: \index{laissez faire}
under unhampered market conditions there is a very realistic tendency
of concentration of wealth by compound interest and cumulative advantage
\index{compound interest} \index{cumulative advantage}  --  again an inescapable dilemma.

\subsubsection{Ambivalent perception of (in)stability}

As Galbraight once mentioned~\cite[Chapter~I]{Galbraith-money}, ``people who are experiencing inflation yearn for stable money and $\ldots$          \index{Galbraith, John Kenneth}
those who are accepting the discipline and the costs of stability come to accept the risks of inflation.
It is this cycle that teaches us that nothing, not even inflation [[or stability]], is [[a]] permanent [[desideratum]].''

\subsection{Summary and outlook}

\begin{quote}
$\ldots$~a blind man eager to see who knows
that the night has no end,
he is still on the go. The rock is still rolling.
\mbox{[[~$\ldots$~]]}
One must imagine Sisyphus happy.
({\em Albert Camus}, in {\it Le Mythe de Sisyphe (English translation: The Myth of Sisyphus)})  \index{Camus, Albert}
\end{quote}

{{S}}{ome} very general options for monetary systems have been enumerated and compared.
The creation of present fiat money {\it via monetization,} as well as its appropriation in various setups has been  discussed.
We have identified private banks as the main source of money through monetization.
Thereby, subject to reserve constraints,
banks absorb (debt related to) assets of value and in exchange issue fiat money in the form of quantity information in deposit money  accounts.

For various reasons discussed earlier, it may not even be possible to determine
the most fundamental entities relevant for money:
\begin{itemize}
\item[$\bullet$]
the volume of money created;
\item[$\bullet$]
the volume of assets and services, both  marketed as well as counterfactual, competing for that money;
\item[$\bullet$]
the price as a measure of value of some asset or service.
\end{itemize}

Thus the value or price is inevitably determined by subjective beliefs and fantasies loosely bound by market constraints.
One may imagine such a monetary system as being ``suspended in thought;'' its continuity,
floating and benign evolution being guaranteed by common faith.

Any such system is vulnerable to crises and business cycles.
For instance, as asset values are subject to disinformation, fraudulent manipulation or hype in anticipation of future profits or losses,
there may be positive and negative feedbacks resulting in price settlements pushing certain equity segments far beyond a stable equilibrium with respect to the rest of  the markets.

Inevitably, the interest levied by banks in return for money created via monetizing debt systematically reallocates
resources toward the financial institutions, and away from industrial and manual production, farming and labor.

In prosperous times the abundance of fiat money guarantees a continuation of economic growth and individual welfare.
Indeed, it is not totally unjustified to argue that the current prosperity of science, public life, and the rise
of living comforts in many countries throughout the globe has been induced by implementing fiat money schemes.
Those stimulus, accompanied by technologies derived from the natural sciences, has created a boom with hitherto unprecedented growth and wealth.

Nevertheless, some very elementary dynamical tendencies almost inevitably  drive economic systems into crises.
One of these tendencies originate from the inevitable subjectivity of economic values,
the greediness of market participants,
and the  positive and negative feedbacks resulting in strong volatility of prices.

Another tendency is
the Matthews Effect \index{Matthews Effect} observed in many other instances~\citep{merton-68}:
it is the {\em cumulative advantage,} the {\em accumulation} of desirable
entities at very few locations, accompanied by a {\em  thinning out} of these entities everywhere else;
sometimes stated as ``the rich getting richer and the poor getting poorer''~\citep{PlutonomyReport2}.

Another common illusion is the belief that Adam Smith's ``invisible hand'' always benignly works for the ``greater good'' of economies;
when in reality economic players are confronted with boundless, relentless greed and prisoner's dilemmas of all sorts.

In view of all the disadvantages of fiat money,
should we even go so far as to abandon the current fiat money system, as well as its associated instruments,
the central bank, and money multiplication by banks within the fractional reserve banking scheme~\citep{ModernMoneyMechanics}?   \index{fractional reserve banking}

Unfortunately, the alternatives appear to be even more troublesome
than the present state of affairs. Any system based on interest-free fiat money creation,
in order to avoid hyperinflation through excessive borrowing associated with ``free'' debt,  has either to rely on unjustifiable privileges or chance.
And any system based on commodity money is heavily dependent on the
quantity of commodities, and also incapable of waging or defending against (economic) war through the effective monetization of future loot or loss.
In effect, we would cripple if not ruin our economies, and would thus behave irresponsibly and destructively,
jeopardizing and ruining our own well-being as well as that of our children.

So, what are the political, economic and social options?
Ought we, for instance, curb banks in their possibilities to create money?
Maybe we should, but if we overdo we cripple our economies by penalizing investments.
If we do not regulate them at all, we stimulate the natural greediness of people,
and foster pyramid scheme type unsustainable business models which assume ever increasing prices (money supply) resulting in economic crises.

The regulatory fine-tuning requires criteria of performance and reliable theories to forecast market behaviors; unfortunately we do not have any such instruments.
But even if such criteria and regulatory instruments will exist in the future  --  which I doubt  --  there might simply be not any possibility to prevent economic crises and the resulting
business cycles.
This may be due to the inherent self-referential character of economic processes, which tend to amplify gains and losses through market hysteria,
and which  --  in a diagonalization type manner~\citep{smullyan-92}  --  are capable of counteracting the very regulatory procedures which are established.

Ought we thus accept occasional monetary crises and the associated business cycles?
I am afraid, yes.

Ought we  accept imbalances of appropriation and a (geometric) redistribution of wealth toward ``the rich,''
and in particular toward the banks and other financial institutions, as well as other aggregates commanding ever increasing amounts of money?
I am afraid, yes. I am unaware of any measure which could counterbalance the accumulation of wealth,
also called the Matthews Effect, in the long run.

There are quite serious political connotations to keep in mind:
As money is the representation of a particular type of asset value, those who control and create money
have equivalent capabilities to deplore economic and political power.
It is quite commonly accepted that societies and empires may be ``steered'' or even dominated by those who have money~\citep{Quigley-TaH};
to the effect that ``money'' renders entire governments; or at least corrupts or overthrows them.
At some point we might wake up and realize that, facilitated by money,  our ``democracies'' have turned into
plutonomies~\citep{PlutonomyReport1,PlutonomyReport2} and oligarchies.

To close this brief discussion in a positive, pragmatic mood, let me mention ways to legally get rich along the monetary lines discussed,
without relying on inherited wealth:
\begin{itemize}
\item[$\bullet$] One of the first and foremost opportunities would be to acquire or start up a central bank if some country would allow one to do so; possibly in exchange of a credit line.
\item[$\bullet$] A fallback option would be to acquire or start up a noncentral bank, or some organization issuing notes which are accepted as some form of exchange payment.
\item[$\bullet$] A third option would be to wait until chance singles one out as a beneficiary of the Matthews Effect. (This may never happen.)
\item[$\bullet$] A fourth option would be to ``ride the tides of inflation and deflation,'' and dynamically increase debt levels,
which must be associated with sustainable levels of interest payments, at proper times.
Without inside information, this, however, may be just as risky as participating in a Ponzi-type pyramid game.
\end{itemize}

On a more existentialist and personal level, I propose to consider money as one of Sisyphus' more absurd assignments~\citep{camus-mos}.


  \addtocontents{toc}{\vspace{\baselineskip}}


%

\end{document}